\documentclass[final,1p,times,12pt]{elsarticle}
\usepackage{amssymb}
\usepackage{amsmath,bm}
\usepackage{colortbl,booktabs}
\usepackage{tabularx}
\usepackage{threeparttable}
\usepackage{multicol,multirow}
\usepackage{subfigure}
\usepackage [font=footnotesize,labelfont=bf]{caption}
\usepackage{rotating}
\usepackage{tikz}
\usepackage{hyperref}
\usepackage{lineno}

\journal{Elsevier}

\begin{document}
\captionsetup[figure]{labelfont={bf},labelformat={default},labelsep=period,name={Fig.}}
\begin{frontmatter}

\title{A generalized non-hourglass updated Lagrangian formulation for SPH solid dynamics}

\author[address1,address2]{Shuaihao Zhang}
\author[address2]{Dong Wu}
\author[address1]{Sérgio D.N. Lourenço}
\author[address2]{Xiangyu Hu \corref{mycorrespondingauthor}}
\cortext[mycorrespondingauthor]{Corresponding author.}
\ead{xiangyu.hu@tum.de}
\address[address1]{Department of Civil Engineering, The University of Hong Kong, Pokfulam, Hong Kong SAR, China}
\address[address2]{School of Engineering and Design, Technical University of Munich, 85748 Garching, Germany}

\begin{abstract}
Hourglass modes, characterized by zigzag particle and stress distributions, are a common numerical instability encountered when simulating solid materials with updated Lagrangian smoother particle hydrodynamics (ULSPH). 
While recent solutions have effectively addressed this issue in elastic materials using an essentially non-hourglass formulation, extending these solutions to plastic materials with more complex constitutive equations has proven challenging due to the need to express shear forces in the form of a velocity Laplacian. 
To address this, a generalized non-hourglass formulation is proposed within the ULSPH framework, suitable for both elastic and plastic materials. 
Specifically, a penalty force is introduced into the momentum equation to resolve the disparity between the linearly predicted and actual velocities of neighboring particle pairs, thereby mitigating the hourglass issue. 
The stability, convergence, and accuracy of the proposed method are validated through a series of classical elastic and plastic cases, with a dual-criterion time-stepping scheme to improve computational efficiency.
The results show that the present method not only matches or even surpasses the performance of the recent essentially non-hourglass formulation in elastic cases but also performs well in plastic scenarios.
\end{abstract}


\end{frontmatter}
%
%
\section{Introduction}
\label{introduction}

The smoothed particle hydrodynamics (SPH) \cite{gingold1977smoothed, lucy1977numerical} is a purely Lagrangian particle-based method where all physical quantities are carried and updated by particles. 
Since its inception in 1977, over several decades, SPH has evolved into a robust numerical method capable of simulating fluid \cite{morris1997modeling, luo2016particle, price2012smoothed}, solid \cite{johnson1996sph, gray2001sph, lee2019total, lee2013development, lee2023entropy}, and fluid-structure interaction \cite{antoci2007numerical, khayyer20213d, khayyer2018enhanced, hwang2014development} problems effectively.
The SPH method can be categorized into updated Lagrangian SPH (ULSPH) \cite{gray2001sph, monaghan2000sph} and total Lagrangian SPH (TLSPH) \cite{vignjevic2006sph}, with ULSPH requiring the update of particle configurations (neighboring particles) at each step, while TLSPH does not necessitate this.
Although TLSPH reduces the computational time required to update particle configurations, ULSPH offers greater advantages in handling material failure and fracture.
When SPH was initially introduced, the ULSPH format was employed and subsequently validated for simulating the motion behavior of both fluids and solids \cite{gingold1977smoothed, lucy1977numerical, morris1997modeling, gray2001sph}.

When simulating solids with ULSPH, there are primarily two types of numerical instabilities.
The first one is known as tensile instability, which was identified by Swegle et al. \cite{swegle1995smoothed} in the standard ULSPH formulation in 1995, manifesting as particle clustering and the presence of non-physical voids.
Subsequently, Monaghan \cite{monaghan2000sph} and Gray et al. \cite{gray2001sph} identified this instability in various solid dynamics scenarios and proposed an artificial stress term, which introduces a repulsive force between particles to prevent clustering, resolving the issue of tensile instability.
Compared to some other methods addressing tensile instability \cite{randles1996smoothed, johnson1996normalized, dyka1997stress, bonet2001remarks, belytschko2002stability, khayyer2023improved}, this approach garnered broader popularity and adoption.
Recent research by Zhang et al. \cite{zhang2024essentially} suggests that the non-physical voids observed in classic elastic cases such as the oscillating beam and colliding rubber rings, as demonstrated in studies like Monaghan \cite{monaghan2000sph} and Gray et al. \cite{gray2001sph}, are actually attributable to hourglass modes rather than tensile instability. 
By addressing hourglass modes, the numerical instabilities in these elastic cases can be eliminated. 
They also note that, just as there are issues with tensile instability when simulating fluids using ULSPH, similar problems persist when simulating highly complex deformation in solid materials with ULSPH.

Hourglass modes \cite{ganzenmuller2015hourglass} are the second type of numerical instability commonly encountered when simulating solid materials using ULSPH, characterized by a zigzag pattern of particles and stress distributions.
The phenomenon of hourglass modes was initially noted in finite element method (FEM) \cite{flanagan1981uniform, jacquotte1984analysis}.
In ULSPH, the root cause of hourglass modes lies in certain deformation modes where, as the particle configuration changes, the velocity gradients remain constant, known as zero-energy modes \cite{dyka1997stress, vignjevic2000treatment, vignjevic2009review}. 
This leads to erroneous stress calculations, resulting in a zigzag pattern, as described in literature \cite{zhang2024essentially, ganzenmuller2015hourglass,wu2023essentially}.
In 2015, Ganzenm{\"u}ller \cite{ganzenmuller2015hourglass} introduced a widely employed method for controlling hourglass modes in solid materials within the TLSPH framework. Subsequently, in 2023, Wu et al.'s study \cite{wu2023essentially} on TLSPH revealed that hourglass instabilities originate from shear forces. They proposed a formulation for elastic materials to address hourglass modes at their root, by drawing insights from the computation of viscous forces in fluid mechanics \cite{morris1997modeling, Monaghan2005SmoothedPH, hu2006multi}.
Building upon this, Zhang et al. \cite{zhang2024essentially} recently introduced the concept of hourglass modes into ULSPH and developed an essentially non-hourglass formulation for elastic material. By decomposing shear forces into a Laplacian form of velocity, they resolved the numerical instabilities in ULSPH when simulating classical elastic problems.
However, due to the complexity of the constitutive relation of plastic materials, it is challenging to establish the relationship between shear forces and the Laplacian form of velocity. Consequently, Zhang et al.'s method \cite{zhang2024essentially} is limited in its applicability to elastic materials exclusively.

Based on the aforementioned discussions, this study proposes to develop a generalized non-hourglass formulation within the ULSPH framework, applicable to both elastic and plastic materials.
Specifically, we introduce a penalty force to eliminate hourglass modes by addressing the misestimation of shear forces in zero-energy modes, based on the inconsistency between the linearly predicted velocity and the actual velocity of neighboring particle pairs.
This penalty force is directly added to the momentum equation to compute the particle's acceleration, thereby not introducing any additional algorithmic complexity.
A dual-criteria time-stepping scheme, originally proposed for fluid simulations \cite{zhang2020dual} and later adapted for solid simulations \cite{zhang2024essentially}, is employed to enhance computational efficiency.

The remainder of this article is organized as follows.
Section \ref{governing-equation} introduces the fundamental theory of elastic and plastic dynamics. The original SPH formulation and the proposed generalized non-hourglass formulation are detailed in Sections \ref{numerical-method} and \ref{hourglass-control-strategy}, respectively. 
In Section \ref{numerical-examples}, a series of benchmark cases for elastic and plastic dynamics are presented to validate the convergence, accuracy, and stability of the proposed method. 
Finally, Section \ref{conclusions} provides the conclusions. 
For further in-depth research, all computational codes used in this study are open-sourced in the SPHinXsys project \cite{zhang2021sphinxsys}, available at \href{https://www.sphinxsys.org}{https://www.sphinxsys.org}.
%
%
\section{Governing equations and constitutive relations}
\label{governing-equation}
\subsection{Elasticity}
\label{elasticity}
In a Lagrangian framework, the governing equations for continuum mechanics involve the conservation of mass and momentum, and can be defined as
\begin{equation}
    \frac{\text{d} \rho }{\text{d} t} = -\mathbf \rho \nabla \cdot \mathbf v
   \label{continuity-equation}
\end{equation}
\begin{equation}
    \frac{\text{d} \mathbf v}{\text{d} t} = \frac{1}{\mathbf \rho}\nabla \cdot \bm{\sigma} + \mathbf g
   \label{momentum-equation}
\end{equation}
where ${\rho}$ is the density, $\mathbf{v}$ is velocity, $\bm{\sigma}$ is the stress tensor, and $\mathbf{g}$ is the body force. 
The total stress tensor, denoted as $\bm{\sigma}$, can be decomposed into two components: hydrostatic pressure $p$ and shear stress $\bm{\sigma}^s$, as illustrated below
\begin{equation}
    \bm{\sigma} = -p \mathbf I + \bm{\sigma}^s
   \label{pressure-shear-stress}
\end{equation}
where $\mathbf{I}$ is the identity matrix. 
The pressure ${p}$ can be determined from the density using an artificial equation of state \cite{gray2001sph}, 
i.e., $p=c_0^2(\rho - \rho_0)$, 
with ${\rho_0}$ and ${\rho}$ being the initial and the current density respectively.
${c_0}$ is the sound speed, and is defined as ${c_0}=\sqrt{{K}/{\rho_0}}$, where $K$ is the bulk modulus.
Taking into account Eq. \eqref{momentum-equation} and Eq. \eqref{pressure-shear-stress}, the acceleration associated with the volumetric component (hydrostatic pressure) and the deviatoric component (shear stress) of the stress tensor can be expressed as follows:
\begin{equation}
    \dot{\mathbf v}^p=-\frac{1}{\mathbf \rho}\nabla p
   \label{normal-acc}
\end{equation}
\begin{equation}
    \dot{\mathbf v}^s=\frac{1}{\mathbf \rho}\nabla \cdot \bm{\sigma}^s
   \label{shear-acc}
\end{equation}
In this case, the velocity change rate (acceleration) caused by hydrostatic pressure and shear stress is represented by $\dot{\mathbf v}^p$ and $\dot{\mathbf v}^s$ respectively. Therefore, the total velocity change rate, denoted as $\dot{\mathbf v}$, can be expressed as the sum of $\dot{\mathbf v}^p$, $\dot{\mathbf v}^s$, and the gravitational acceleration $\mathbf{g}$.
The shear stress $\bm{\sigma}^s$ is obtained by integrating the rate of shear stress with respect to time.
\begin{equation}
    \bm{\sigma}^s=\int_{0}^{t} \dot{\bm{\sigma}}^s  \text{d}t 
   \label{shear-stress-integral}
\end{equation}
In the case of a linear elastic model, the rate of shear stress $\dot{\bm{\sigma}}^s$ is defined as follows
\begin{equation}
    \dot{\bm{\sigma}}^s = 2G{\dot{\bm \varepsilon}^s}
   \label{stress-rate}
\end{equation}
where $G$ is the shear modulus and ${d}$ represents the space dimension. 
${\dot{\bm \varepsilon}^s}$ is deviatoric strain rate and ${\dot{\bm \varepsilon}^s} = \dot{\bm \varepsilon}  -\frac{1}{d}tr(\dot{\bm \varepsilon})\mathbf I$. ${tr(\dot{\bm \varepsilon})}$ is the trace of strain rate $\dot{\bm \varepsilon}$, and $d=2$ or $3$ is the space dimension.
The strain rate is defined as
\begin{equation}
    \dot{\bm \varepsilon} =\frac{1}{2} \left( \nabla \mathbf v + (\nabla \mathbf v)^T \right)
   \label{strain-rate}
\end{equation}
Here, ${\nabla \mathbf v}$ represents the velocity gradient, and the superscript $T$ denotes the transpose of a tensor.
\subsection{Plasticity}
\label{plasticity}
The $J_2$ plasticity model \cite{borja2013plasticity} is adopted in this study.
The yield function is expressed as
\begin{equation}
   f(J_2, \alpha )=\sqrt{2J_2} - \sqrt{\frac{2}{3}}(\kappa \alpha +{\sigma}_Y)
  \label{yield-criterion}
\end{equation}
where $J_2=\frac{1}{2}{\bm{\sigma}}^s:{\bm{\sigma}}^s$ is the second invariant of stress tensor. 
$\kappa$ is the hardening modules, and $\alpha$ is the hardening factor.
$\sigma_Y$ is the initial flow stress, which is also called the yield stress. 
Refer to \cite{chaves2013notes}, the shear stress rate for the $J_2$ plasticity model can be expressed as
\begin{equation}
	\dot{\bm{\sigma}}^s = 2G{\dot{\bm \varepsilon}^s}-\dot{\lambda} \frac{\sqrt{2}G}{\sqrt{J_2}}\bm{\sigma}^s
   \label{J2-stress-rate}
\end{equation}
Here, $\lambda$ is the plastic multiplier, and its change rate $\dot{\lambda}$ is defined as \cite{chaves2013notes}
\begin{equation}
	\dot{\lambda}=\frac{\bm{\sigma}^s:\dot{\bm{\varepsilon}}}{(1+\kappa/3G){\sqrt{2J_2}}}
   \label{plastic-multiplier-rate}
\end{equation}
If the stress state exceeds the yield surface ($f>0$), further operations are necessary to bring the stress state back onto the yield surface. This process, known as the stress return mapping algorithm \cite{borja2013plasticity, simo2006computational}, involves adjusting the stress state to satisfy the yield condition.
The relationship between the shear stress $\widetilde{\bm{\sigma}}^s$ after the return mapping and the shear stress ${\bm{\sigma}}^s$ before the return mapping can be expressed as follows \cite{borja2013plasticity, simo2006computational}
\begin{equation}
	\widetilde{\bm{\sigma}}^s=\frac{\kappa\alpha +{\sigma}_Y}{\sqrt{3J_2}}{\bm{\sigma}}^s
   \label{return-mapping}
\end{equation}
The acceleration induced by shear stress for plastic material can be described as
\begin{equation}
   \dot{\mathbf v}^s=\frac{1}{\mathbf \rho}\nabla \cdot \widetilde{\bm{\sigma}}^s
  \label{shear-acc-J2}
\end{equation}
%
%
\section{Numerical method}
\label{numerical-method}

\subsection{SPH discretization}
\label{SPH-discretization}
The continuity equation is discretized as 
\begin{equation}
   \frac{\text{d} \rho_i }{\text{d} t} = \rho_i \sum_{j} \mathbf v_{ij} {\nabla_i W_{ij}} V_j
  \label{continuity-equation-discrete}
\end{equation}
We employ a Riemann solver \cite{zhang2017weakly, zhang2024riemann} to discretize the momentum equation for the hydrostatic pressure component
\begin{equation}
    \frac{\text{d} \mathbf v_i^p}{\text{d} t} = -2 \frac{1}{\rho_i}\sum_{j} P^* {\nabla_i W_{ij}} V_j
   \label{normal-accelaration-discrete}
\end{equation}
In this study, the variables and notations used are as follows: $W_{ij}$ represents the kernel function $W({\mathbf r}_i - {\mathbf r}_j, h)$, where ${\mathbf r}$ denotes the position of a particle and ${h}$ represents the smoothing length. 
The subscripts ${i}$ and ${j}$ refer to particle numbers, and $V_j$ is the volume of particle $j$.
The unit vector pointing from particle ${j}$ to particle ${i}$ is denoted as $\mathbf e_{ij}$ and $\mathbf e_{ij}=\mathbf e_{i}-\mathbf e_{j}$. 
$\mathbf v_{ij} = \mathbf v_i - \mathbf v_j$ represents the relative velocity between the two neighbor particles.
The derivative of the kernel function is given by ${\nabla_i W_{ij}} = \frac{\partial W({r}_{ij}, h)}{\partial {r}_{ij}} \mathbf e_{ij}$, where ${r}_{ij} = |\mathbf r_{i} - \mathbf r_{j}|$ represents the distance between the two particles.
The quantities $P^*$, which are obtained from the Riemann solver \cite{zhang2017weakly}, correspond to the solutions of an inter-particle Riemann problem along the unit vector pointing from particle $i$ to particle $j$.
$P^*$ are defined as \cite{zhang2017weakly}
\begin{equation}
    P^* = \frac{\rho_L c_L P_R+\rho_R c_R P_L+ \rho_L c_L \rho_R c_R (U_L-U_R)}{\rho_L c_L + \rho_R c_R}
   \label{P-Riemann}
\end{equation}
The subscripts $L$ and $R$ denote the left and right states of the Riemann problem, respectively, and they are defined as follows
\begin{equation}
	\begin{cases}
		(\rho_L, U_L, P_L, c_L ) = (\rho_i, \mathbf v_i \cdot \mathbf e_{ij}, P_i, c_{0i}) \\
		(\rho_R, U_R, P_R, c_R ) = (\rho_j, \mathbf v_j \cdot \mathbf e_{ij}, P_j, c_{0j}) 
	\end{cases}
   \label{left-right-states}
\end{equation}
For elastic dynamics, the acceleration $\dot{\mathbf v}^s$ related to shear stress can be discretized by
\begin{equation}
    \frac{\text{d} \mathbf v_i^s}{\text{d} t} = \frac{1}{\rho_i} \sum_{j} \left ( \bm{\sigma}^s_i + \bm{\sigma}^s_j \right ) \cdot{\nabla_i W_{ij}} V_j
   \label{shear-accelaration-discrete}
\end{equation}
While for plastic behavior, it should be written as
\begin{equation}
   \frac{\text{d} \mathbf v_i^s}{\text{d} t} = \frac{1}{\rho_i}
   \sum_{j} \left ( \widetilde{\bm{\sigma}}^s_i + \widetilde{\bm{\sigma}}^s_j \right ) \cdot{\nabla_i W_{ij}} V_j
  \label{shear-accelaration-discrete-J2}
\end{equation}
The velocity gradient in Eq. \eqref{strain-rate} can be discretized as \cite{espanol2003smoothed}
\begin{equation}
   {\nabla \mathbf v}=\sum_{j} \mathbf v_{ij} \otimes \left(\mathbf B_i {\nabla_i W_{ij}} \right) V_j
  \label{velocity-gradient-discrete-correction}
\end{equation}
$\mathbf B_i$ is the correction matrix for kernel gradient \cite{randles1996smoothed, bonet2002simplified, ren2023efficient} and is defined as
\begin{equation}
    \mathbf B_i = - \left({\sum_{j} \mathbf r_{ij} \otimes {\nabla_i W_{ij}}  V_j} \right)^{-1}
   \label{correction-matrix}
\end{equation}
\subsection{Time integration scheme}
\label{time-integration}
The dual-criteria time stepping strategy \cite{zhang2024essentially, zhang2020dual}, which employs a larger advection time step $\bigtriangleup t_{ad}$ and a smaller acoustic time step $\bigtriangleup t_{ac}$, is adopted in this study to enhance the computational efficiency.
The particle configuration is updated during the advection time step $\bigtriangleup t_{ad}$, which can be defined as
\begin{equation}
    \bigtriangleup t_{ad} = CFL_{ad}\frac{h}{\left\lvert \mathbf v \right\rvert_{max} }
   \label{advection-time-step}
\end{equation}
The advection time step $\bigtriangleup t_{ad}$ is determined based on the Courant-Friedrichs-Lewy (CFL) condition, where $CFL_{ad}=0.2$. It is calculated using the maximum particle advection speed $\left\lvert \mathbf v \right\rvert_{max}$ and the smoothing length $h$.
The acoustic time step $\bigtriangleup t_{ac}$, which governs the update of particle properties such as velocity and density, is given by
\begin{equation}
    \bigtriangleup t_{ac} = CFL_{ac}\frac{h}{c_0+\left\lvert \mathbf v \right\rvert_{max} }
   \label{acoustic-time-step}
\end{equation}
where $CFL_{ac}=0.4$ and $c_0$ is the sound speed. 

Next, the position-based Verlet scheme \cite{zhang2021multi} is employed for the acoustic time integration.
During the acoustic time step, the beginning of which is denoted by the superscript $n$, the midpoint is represented by the superscript $n+\frac{1}{2}$, and the new time step is indicated by the superscript $n+1$. In the Verlet scheme, the particle position and density are initially updated to the midpoint using the following procedure
\begin{equation}
	\begin{cases}
		{\mathbf r}^{n+\frac{1}{2}}={\mathbf r}^n+ \frac{1}{2} {\bigtriangleup t_{ac}} {\mathbf v}^n  \\
		{\rho}^{n+\frac{1}{2}}={\rho}^n+ \frac{1}{2} {\bigtriangleup t_{ac}} \left({\frac{\text{d} \rho}{\text{d} t}} \right)^n        \\
	\end{cases}
   \label{time-step-half}
\end{equation}
Subsequently, once the particle acceleration is determined, the velocity is updated to the new time step.
\begin{equation}
    {\mathbf v}_{n+1}={\mathbf v}_{n} + {\bigtriangleup t_{ac}} \left({\frac{\text{d} \mathbf v}{\text{d} t}} \right)^n
   \label{time-step-velocity}
\end{equation}
Finally, the particle position and density are updated to the new time step using the following process.
\begin{equation}
	\begin{cases}
		{\mathbf r}^{n+1}={\mathbf r}^{n+\frac{1}{2}}+ \frac{1}{2} {\bigtriangleup t_{ac}} {\mathbf v}^{n+1}  \\
		{\rho}^{n+1}={\rho}^{n+\frac{1}{2}}+ \frac{1}{2} {\bigtriangleup t_{ac}} \left({\frac{\text{d} \rho}{\text{d} t}} \right)^{n+1}        \\
	\end{cases}
   \label{time-step-full}
\end{equation}
%
%
\section{Generalized non-hourglass formulation}
\label{hourglass-control-strategy}
Petschek and Hanson \cite{petschek1968difference} were the first to recognize that the presence of hourglass modes can be attributed to the absence of bilinear terms in the velocity field in the finite difference method. 
This observation was further substantiated by Belytschko \cite{belytschko1974finite} within the framework of finite element analysis.
SPH exhibits certain similarities to the mean stress-strain description of a one-integration point finite element \cite{flanagan1981uniform}.
The kernel approximation in SPH results in a smeared-out representation of field variables defined at the center of SPH particles \cite{ganzenmuller2015hourglass}.
Although the field variables may vary throughout the simulation domain, each particle assumes a constant or mean field value within its local neighborhood. 
This characteristic of SPH is directly connected to the nodal integration approach, which employs a piecewise constant integration technique \cite{ganzenmuller2015hourglass}.
Hourglass modes are characterized by nodal displacements or velocities that are incompatible with the linear field.

In TLSPH, a penalty force \cite{ganzenmuller2015hourglass} is introduced to eliminate hourglass modes by addressing the discrepancy between the actual displacement values and their linear estimates. 
Such an approach ensures the accuracy and stability of the simulation results.
While in ULSPH, the displacement can be obtained by integrating the velocity with respect to time.
Firstly, the velocity difference between point $i$ and point $j$, obtained through linear prediction, is given by
\begin{equation}
   {\mathbf v}_{ij}^{linear}=(\nabla {\mathbf v}_{i} + \nabla {\mathbf v}_{j})\cdot \mathbf r_{ij}
  \label{vij_linear}
\end{equation}
There exists a difference, denoted as $\widehat{\mathbf v}_{ij}$, between the velocity difference ${\mathbf v}_{ij}^{linear}$ obtained through linear prediction and the actual velocity difference ${\mathbf v}_{ij}$.
\begin{equation}
   \widehat{\mathbf v}_{ij}={\mathbf v}_{ij}-{\mathbf v}_{ij}^{linear}={\mathbf v}_{ij}- \left(\nabla {\mathbf v}_{i} + \nabla {\mathbf v}_{j} \right)\cdot \mathbf r_{ij}
  \label{vij_error}
\end{equation}
The penalty force $\widehat{\mathbf f}_{ij}$ between particle $i$ and particle $j$, which is used to eliminate hourglass modes, can be obtained by integrating the $\widehat{\mathbf v}_{ij}$ over time.
\begin{equation}
   \widehat{\mathbf f}_{ij}=\xi G 
   \int_{0}^{t} {\frac{\widehat{\mathbf v}_{ij}}{\left\lvert \mathbf r_{ij} \right\rvert}} 
   {\frac{\partial  W_{ij}}{\partial {r}_{ij}} V_i V_j} 
   \text{d}t
  \label{penalty-force}
\end{equation}
where $\xi$ is a positive coefficient and needs to be calibrated by numerical experiments.
By substituting Eq. \eqref{penalty-force} into Eq. \eqref{shear-accelaration-discrete}, we can derive a formulation of shear acceleration that eliminates hourglass modes for elastic dynamics.
\begin{equation}
   \frac{\text{d} \mathbf v_i^s}{\text{d} t} = \frac{1}{\rho_i}
   \sum_{j} \left[ \left(\bm{\sigma}^s_i + \bm{\sigma}^s_j \right) \cdot \mathbf e_{ij}  
   \frac{\partial  W_{ij}}{\partial {r}_{ij}} V_j + 
   {\xi G\int_{0}^{t} {\frac{\widehat{\mathbf v}_{ij}}{\left\lvert \mathbf r_{ij} \right\rvert} \frac{\partial  W_{ij}}{\partial {r}_{ij}} V_j}  
   \text{d}t}  \right]
  \label{shear-accelaration-discrete-elastic}
\end{equation}
By examining Eq. \eqref{vij_linear} to Eq. \eqref{penalty-force}, we can observe that the penalty force, aimed at eliminating hourglass modes, arises from the disparity between the linearly predicted and actual velocities of neighboring particle pairs. 
This implies that, within the particle's support domain, the deformation field is constrained to exhibit local linearity \cite{ganzenmuller2015hourglass}. 
Consequently, such an hourglass control method hampers nonlinear deformations, including plastic deformation \cite{ganzenmuller2015hourglass}. Given this analysis, it is necessary to reduce the penalty force specifically for plastic deformation \cite{ganzenmuller2015hourglass}. 

First, we will perform some transformations on the expression for the shear acceleration in elastic materials, and Eq. \eqref{shear-accelaration-discrete-elastic} is rewritten as
\begin{equation}
   \frac{\text{d} \mathbf v_i^s}{\text{d} t} = \frac{1}{\rho_i}
   \sum_{j} \left[ \left( {\bm{\sigma}}^s_i + {\bm{\sigma}}^s_j \right) \cdot \mathbf e_{ij}  
   \frac{\partial  W_{ij}}{\partial {r}_{ij}} V_j + 
   {\xi G \int_{0}^{t} { \overline{\bm \varphi}_{ij} \frac{\widehat{\mathbf v}_{ij}}{\left\lvert \mathbf r_{ij} \right\rvert} \frac{\partial  W_{ij}}{\partial {r}_{ij}} V_j}  
   \text{d}t}  \right]
  \label{shear-accelaration-discrete-adjust}
\end{equation}
where $\overline{\bm \varphi}_{ij}=(\bm \varphi_i+\bm \varphi_j)/2$ and $\bm \varphi_i$ is defined as
\begin{equation}
	\bm \varphi_i = \bm{\bm{\sigma}}^{s}_i \bm{\bm{\sigma}}^{-s}_i = \mathbf{I}
   \label{scale-mateix-elastic}
 \end{equation}
where ${\bm{\sigma}}^{s}_i$ is the shear stress and $ \bm{\bm{\sigma}}^{-s}_i$ is the inverse of shear stress.
Hence, $\bm \varphi_{ij}=\mathbf{I}$, indicating the equivalence of Eq. \eqref{shear-accelaration-discrete-adjust} and Eq. \eqref{shear-accelaration-discrete-elastic}.

For plasticity, $\bm \varphi_{ij}$ should be reduced to decrease the penalty force. By borrowing the concept of stress return mapping in plastic materials (Eq. \eqref{return-mapping}), $\bm \varphi_{i}$ in the plastic deformation can be defined as
\begin{equation}
   \bm \varphi_i = {\widetilde{\bm{\sigma}}^s}_i \bm{\bm{\sigma}}^{-s}_i  = \gamma_i \bm{\bm{\sigma}}^{s}_i \bm{\bm{\sigma}}^{-s}_i = \gamma_i \mathbf{I}
  \label{scale-mateix-plastic}
\end{equation}
where $\gamma $ is the scale coefficient in the stress return mapping (Eq. \eqref{return-mapping}) and is expressed as 
\begin{equation}
	\gamma =
	\begin{cases}
		\frac{\kappa\alpha +{\sigma}_Y}{\sqrt{3J_2}}, \ \text{if} \ f(J_2, \alpha ) >0 \\
		1, \ \text{if} \ f(J_2, \alpha ) \leqslant 0        \\
	\end{cases}
   \label{scale-coefficient}
\end{equation}
Finally, the formulation without hourglass modes in the plastic case can be obtained as
\begin{equation}
   \frac{\text{d} \mathbf v_i^s}{\text{d} t} = \frac{1}{\rho_i}
   \sum_{j} \left[ \left( \widetilde{\bm{\sigma}}^s_i + \widetilde{\bm{\sigma}}^s_j \right) \cdot \mathbf e_{ij}  
   \frac{\partial  W_{ij}}{\partial {r}_{ij}} V_j + 
   {\xi G \int_{0}^{t} { 
	\overline{\gamma}_{ij}
	\frac{\widehat{\mathbf v}_{ij}}{\left\lvert \mathbf r_{ij} \right\rvert} \frac{\partial  W_{ij}}{\partial {r}_{ij}} V_j}  
   \text{d}t}  \right]
  \label{shear-accelaration-discrete-plastic}
\end{equation}
where $\overline{\gamma}_{ij} = ( \gamma_i + \gamma_j )/2$.
For elasticity, $\overline{\gamma}_{ij} = 1$, and thus Eq. \eqref{shear-accelaration-discrete-plastic} is same with Eq. \eqref{shear-accelaration-discrete-elastic}.
In the context of plastic deformation, $\overline{\gamma}_{ij}$ is less than 1, indicating that the penalty force is reduced. 
Based on our numerical experiments, $\xi $ is generally set to 4 for elasticity and 0.2 for plasticity, eliminating the need for tuning for each case.
%
%
\section{Numerical examples}
\label{numerical-examples}

In this section, we evaluate several benchmark cases, comparing the results with analytical solutions and those from previous numerical studies, both qualitatively and quantitatively. 
To assess the accuracy, stability, and robustness of the proposed method, we contrast our findings with those obtained using the original ULSPH formulation, the ULSPH formulation with artificial stress \cite{gray2001sph}, and the recent essentially non-hourglass ULSPH formulation \cite{zhang2024essentially}. 
For clarity, we define several abbreviations for different SPH methods used throughout the article: "SPH-OG" for the original ULSPH method, "SPH-OAS" for the original ULSPH with artificial stress \cite{gray2001sph}, "SPH-ENOG" for the essentially non-hourglass ULSPH formulation \cite{zhang2024essentially}, and "SPH-GNOG" for the present generalized non-hourglass ULSPH formulation.

In this study, we use the 5th-order Wendland kernel \cite{wendland1995piecewise} with a smoothing length of $h = 1.3dp$
and a cut-off radius of 2.6$dp$, where $dp$ is the initial particle spacing, for all cases.
All physical quantities in this article are presented in dimensionless form.
\subsection{2D oscillating plate}
\label{2D-oscillating-plate}
\begin{figure}[htb!]
	\centering
	\includegraphics[trim = 0cm 0cm 0cm 0cm, clip,width=.85\textwidth]{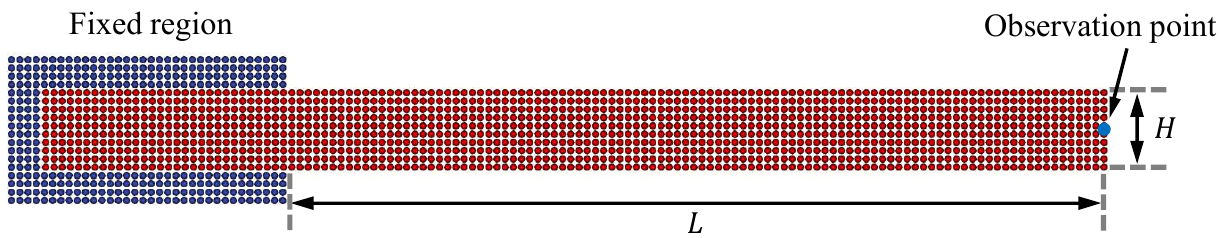}
	\caption{2D oscillating plate: model setup.}
	\label{figs:2D-plate-setup}
\end{figure}

In order to verify the proposed method in elastic dynamics, a 2D plate with one fixed edge, as depicted in Fig. \ref{figs:2D-plate-setup}, is employed. The obtained results are then compared with both previous theoretical \cite{landau2013course} and numerical \cite{gray2001sph, zhang2024essentially} solutions. The plate has a length $L$ and a thickness $H$, with the left part being fixed to create a cantilever configuration.

For analysis, an observation point is positioned at the midpoint of the tail to measure the vertical displacement, defined as the deflection. A positive value indicates upward displacement, while a negative value represents downward displacement.
Additionally, an initial velocity $v_y$, perpendicular to the plate strip, is applied to the system.
\begin{equation}
   v_y(x) = v_f c_0 \frac{f(x)}{f(L)}
  \label{2D-plate-initial-velocity}
\end{equation}
where
\begin{equation}
   \begin{aligned}
       f(x)=(\sin (kL) + \sinh(kL))(\cos(kx)-\cosh(kx)) \\
       - (\cos(kL)+\cosh(kL))(\sin(kx)-\sinh(kx))
   \end{aligned}
  \label{2D-plate-fx}
\end{equation}
Here, $v_f$ is an input parameter and $c_0$ is the sound speed.
$kL=1.875$ is determined by $\cos(kL) \cosh(kL)=-1$. 
The frequency $\omega $ of the oscillating plate is theoretically given by
\begin{equation}
   {\omega}^2=\frac{EH^2k^4}{12\rho_0 (1-{\nu}^4 )}
  \label{2D-plate-frequency}
\end{equation}
where $E$ is the Young's modulus and $\nu$ is the Poisson's ratio.
The material and dimensional parameters in this case are adopted from previous studies \cite{gray2001sph, zhang2017generalized}. Specifically, the density is set to $\rho_0=1000$, Young's modulus is $E=2\times 10^6$, Poisson's ratio is $\nu = 0.3975$, and the plate dimensions are $L=0.2$ and $H=0.02$.

\begin{figure}[htb!]
	\centering
	\includegraphics[trim = 0cm 0cm 0cm 0cm, clip,width=.95\textwidth]{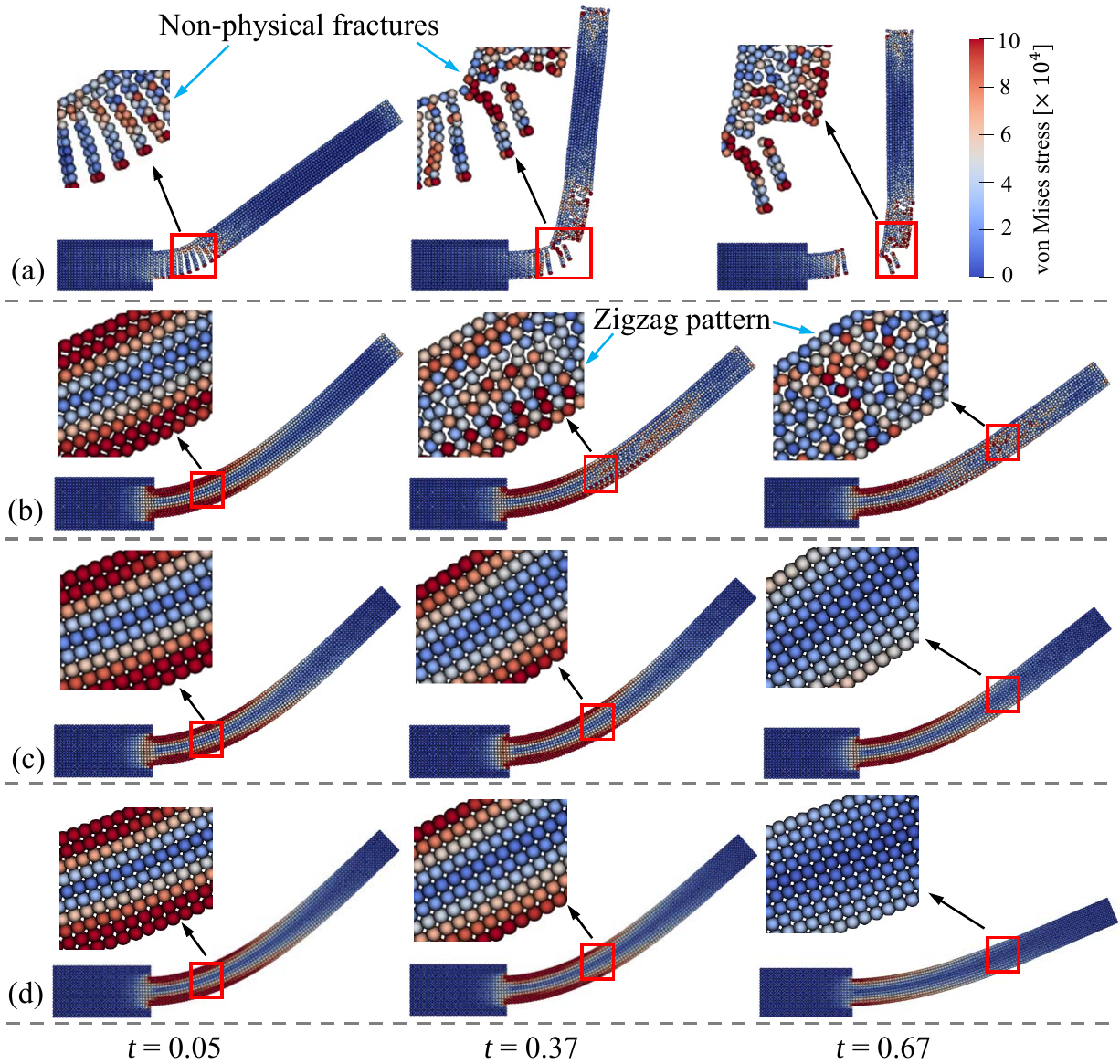}
	\caption{2D oscillating plate: evolution of particle configuration with time (t=0.05, 0.37 and 0.67) for (a) SPH-OG \cite{zhang2024essentially}, (b) SPH-OAS \cite{gray2001sph, zhang2024essentially}, (c) SPH-ENOG \cite{zhang2024essentially}, and (d) SPH-GNOG. Here, $v_f=0.05$, $L=0.2$, and $H=0.02$. The particles are colored by von Mises stress.}
	\label{figs:2D-plate-compare}
\end{figure}

Fig. \ref{figs:2D-plate-compare} illustrates snapshots of the results obtained at different time instances when simulating the 2D oscillating plate using various numerical methods.
When simulating elastic deformation using SPH-OG, as shown in Fig. \ref{figs:2D-plate-compare}a, significant numerical instabilities including non-physical fractures and zigzag pattern can be observed.
These instabilities manifest themselves at the initial stage of the simulation ($t = 0.05$), resulting in a non-uniform distribution of particles and a distorted profile of von Mises stress, indicating the presence of hourglass modes.
Fig. \ref{figs:2D-plate-compare}b presents the outcomes obtained through SPH-OAS, which effectively mitigates the occurrence of non-physical fractures. 
However, the persistence of zigzag patterns becomes visually evident at $t = 0.37$ due to the integral nature of the error in the original formulation, resulting in its gradual accumulation over time.
Similar to SPH-ENOG (Fig. \ref{figs:2D-plate-compare}c), the present SPH-GNOG (Fig. \ref{figs:2D-plate-compare}d) yields results devoid of non-physical fractures and zigzag patterns even at $t = 0.67$. The particle distribution remains uniform, and the stress profile exhibits smoothness.

\begin{figure}[htb!]
	\centering
	\includegraphics[trim = 0cm 0cm 0cm 0cm, clip,width=.6\textwidth]{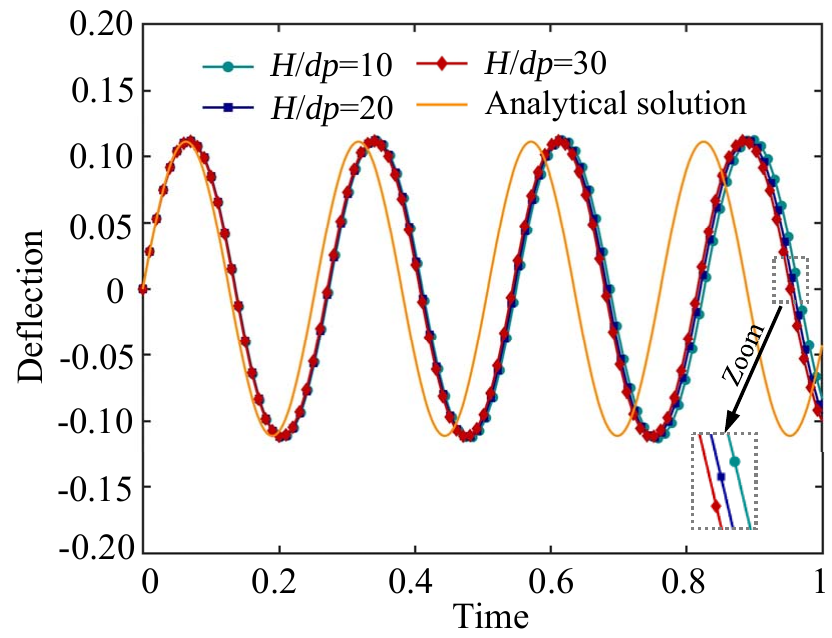}
	\caption{2D oscillating plate: temporal evolution of deflection at various resolutions. Here, $v_f=0.05$, $L=0.2$, and $H=0.02$.}
	\label{figs:2D-plate-deflection}
\end{figure}

\begin{figure}[htb!]
	\centering
	\includegraphics[trim = 0cm 0cm 0cm 0cm, clip,width=.7\textwidth]{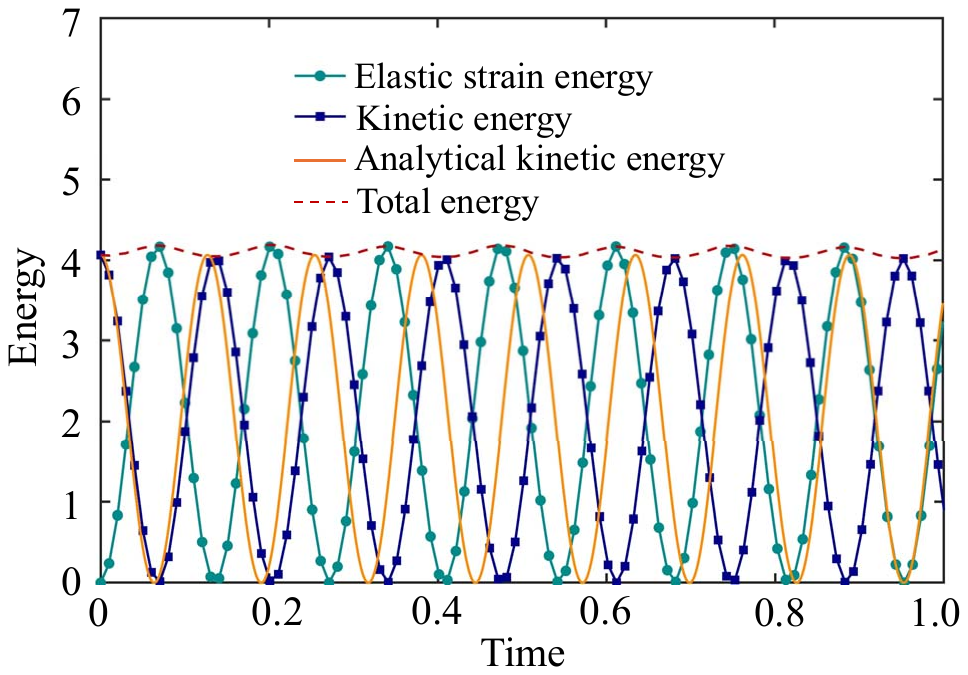}
	\caption{2D oscillating plate: temporal evolution of elastic strain energy, kinetic energy, and total energy. Here, $v_f=0.05$, $L=0.2$, and $H=0.02$.}
	\label{figs:2D-plate-energy}
\end{figure}

In Fig. \ref{figs:2D-plate-deflection}, the convergence of the new formulation is validated. Three cases with varying resolutions ($H/dp=10$, $H/dp=20$, and $H/dp=30$) are tested, and the time-dependent variations in deflection is shown in Fig. \ref{figs:2D-plate-deflection}. As the resolution increases, the discrepancies between solutions diminish, consistent with results from the literature \cite{gray2001sph, zhang2017generalized, wu2023essentially, khayyer2018enhanced}, indicating the convergence of the proposed algorithm. The theoretical deflection values over time are also plotted in Fig. \ref{figs:2D-plate-deflection} \cite{khayyer2018enhanced}. It is evident that higher resolutions lead to numerical results converging toward the theoretical values.
Fig. \ref{figs:2D-plate-energy} illustrates the time-dependent variations of elastic strain energy, kinetic energy, and total energy for the present SPH-ENOG with $H/dp=30$. The theoretical solution for kinetic energy is also included for comparison. The results show that the computed kinetic energy closely aligns with the theoretical values. Additionally, kinetic energy and elastic strain energy fluctuate alternately, while their sum, the total energy, remains nearly constant \cite{zhang2024essentially}.

\begin{figure}[htb!]
	\centering
	\includegraphics[trim = 0cm 0cm 0cm 0cm, clip,width=.9\textwidth]{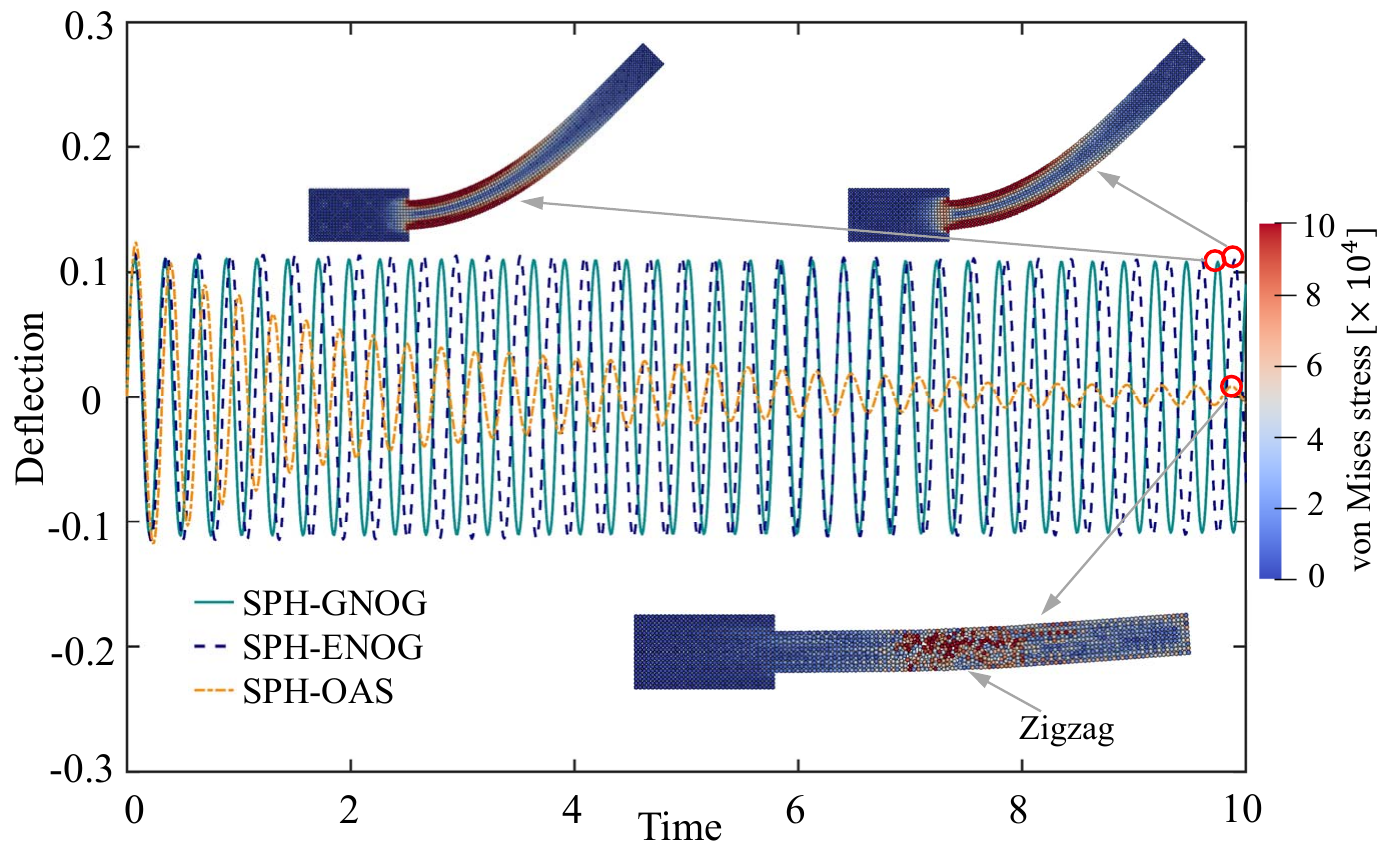}
	\caption{2D oscillating plate: test the long-time stability of the SPH-GNOG. The result is compared with those obtained by SPH-ENOG \cite{zhang2024essentially} and SPH-OAS \cite{zhang2024essentially}. Here, $L$=0.2, $H$=0.02, $H/dp=10$ and $\mathbf v_f$=0.05. The particles are colored by von Mises stress.}
	\label{figs:2D-plate-stress-test}
\end{figure}

Next, we conducted a stress test with a long-time simulation to evaluate the stability of the current algorithm. As shown in Fig. \ref{figs:2D-plate-stress-test}, the simulation lasted for over 30 oscillations, and results from SPH-ENOG and SPH-OAS were included for comparison. All three simulations used a single time step to minimize accumulated integration error in long-time simulations. Fig. \ref{figs:2D-plate-stress-test} also shows the particle distributions obtained using different methods at around $t \approx 10$. Similar to SPH-ENOG, the proposed SPH-GNOG maintained uniform particle and stress distributions until the end of the simulation, while SPH-OAS showed significant hourglass issues at $t \approx 10$. Additionally, with SPH-GNOG, the deflection at $t = 10$ decreased only slightly compared to $t = 0$, due to numerical dissipation introduced by the Riemann solver \cite{zhang2017weakly}. In contrast, SPH-OAS exhibited rapid energy decay, making it unsuitable for long-duration computations.

Furthermore, the accuracy of the algorithm is verified. Table \ref{2D-plate-error} presents the calculated and theoretical values for the first oscillation period of the 2D oscillating plate at different initial velocities. Results obtained using SPH-ENOG and SPH-OAS are also shown for comparison. It can be seen that the average error of the results obtained using SPH-GNOG is $7.5 \%$ compared to the theoretical values, which is on par with the errors of other methods. This demonstrates the accuracy of the proposed SPH-GNOG.
\begin{table}
	\scriptsize
	\centering
	\caption{2D oscillating plate: comparison of the first oscillation period $T$ obtained from the present SPH-GNOG, SPH-OAS \cite{gray2001sph}, SPH-ENOG \cite{zhang2024essentially} and analytical solutions. Here, $L$=0.2, $H$=0.02 and $H/dp=30$.}
	\begin{tabularx}{8.5cm}{@{\extracolsep{\fill}}lcccc}
		\hline
		$v_f$ & 0.001 & 0.01 & 0.03 & 0.05\\
		\hline
        $T$ (Analytical) & 0.254 & 0.254 & 0.254 & 0.254  \\
		$T$ (SPH-GNOG) & 0.275 & 0.273 & 0.272 & 0.272 \\
		$T$ (SPH-OAS) & 0.273 & 0.273 & 0.275 & 0.278 \\
      $T$ (SPH-ENOG) & 0.262 & 0.263 & 0.268 & 0.279 \\
        Error (SPH-GNOG) & 8.3$\%$ & 7.5$\%$ & 7.1$\%$ & 7.1$\%$ \\
		Error (SPH-OAS) & 7.5$\%$ & 7.5$\%$ & 8.3$\%$ & 9.4$\%$ \\
      Error (SPH-ENOG) & 3.1$\%$ & 3.5$\%$ & 5.5$\%$ & 9.8$\%$ \\
		\hline
	\end{tabularx}
	\label{2D-plate-error}
\end{table}
\subsection{Bending column}
\label{bending-column}
\begin{figure}[htb!]
	\centering
	\includegraphics[trim = 0cm 0cm 0cm 0cm, clip,width=.25\textwidth]{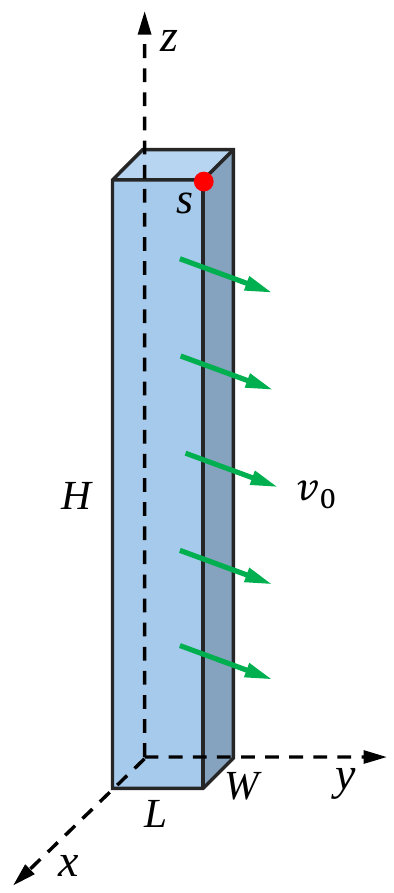}
	\caption{3D bending column: model setup.}
	\label{figs:bending-column-setup}
\end{figure}

In this section, we test the stability and accuracy of the proposed algorithm in 3D elastic dynamics. As shown in Fig. \ref{figs:bending-column-setup}, a rubber-like column with a length and width of $L=W=1$ and a height of $H=6$, fixed at the base, begins to move under an initial velocity $v_0=10 \left (\frac{\sqrt{3}}{2},\frac{1}{2},0 \right )^T$\cite{aguirre2014vertex}. 
In this bending-dominated case, significant tensile forces develop on the outer side of the column. 
An elastic constitutive model (section \ref{elasticity}) is applied with density $\rho_0=1100$, Young's modulus $E=2\times 1.7^7$, and Poisson's ratio $\nu = 0.45$.
The coordinates of the observation point $s$ are $\left (1,1,6 \right )$.

\begin{figure}[htb!]
	\centering
	\includegraphics[trim = 0cm 0cm 0cm 0cm, clip,width=1.0\textwidth]{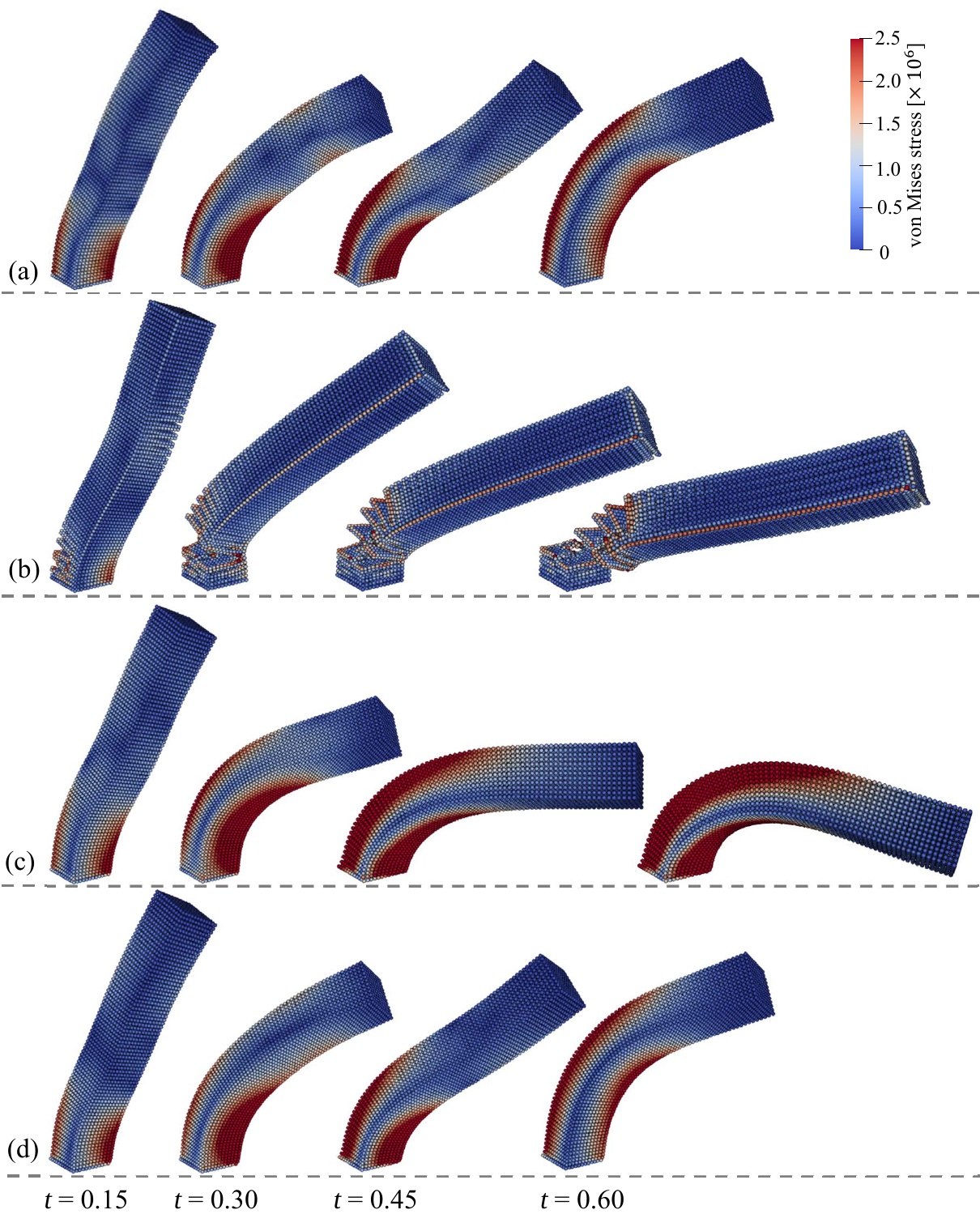}
	\caption{3D bending column: evolution of particle configuration with time (t=0.15, 0.30, 0.45 and 0.60) for (a) TLSPH \cite{wu2023essentially}, (b) SPH-OG, (c) SPH-ENOG \cite{zhang2024essentially}, and (d) SPH-GNOG. Here, $L/dp=12$ and $dp$ is the initial particle spacing. The particles are colored by von Mises stress.}
	\label{figs:bending-column-stress}
\end{figure}

\begin{figure}[htb!]
	\centering
	\includegraphics[trim = 0cm 0cm 0cm 0cm, clip,width=0.9\textwidth]{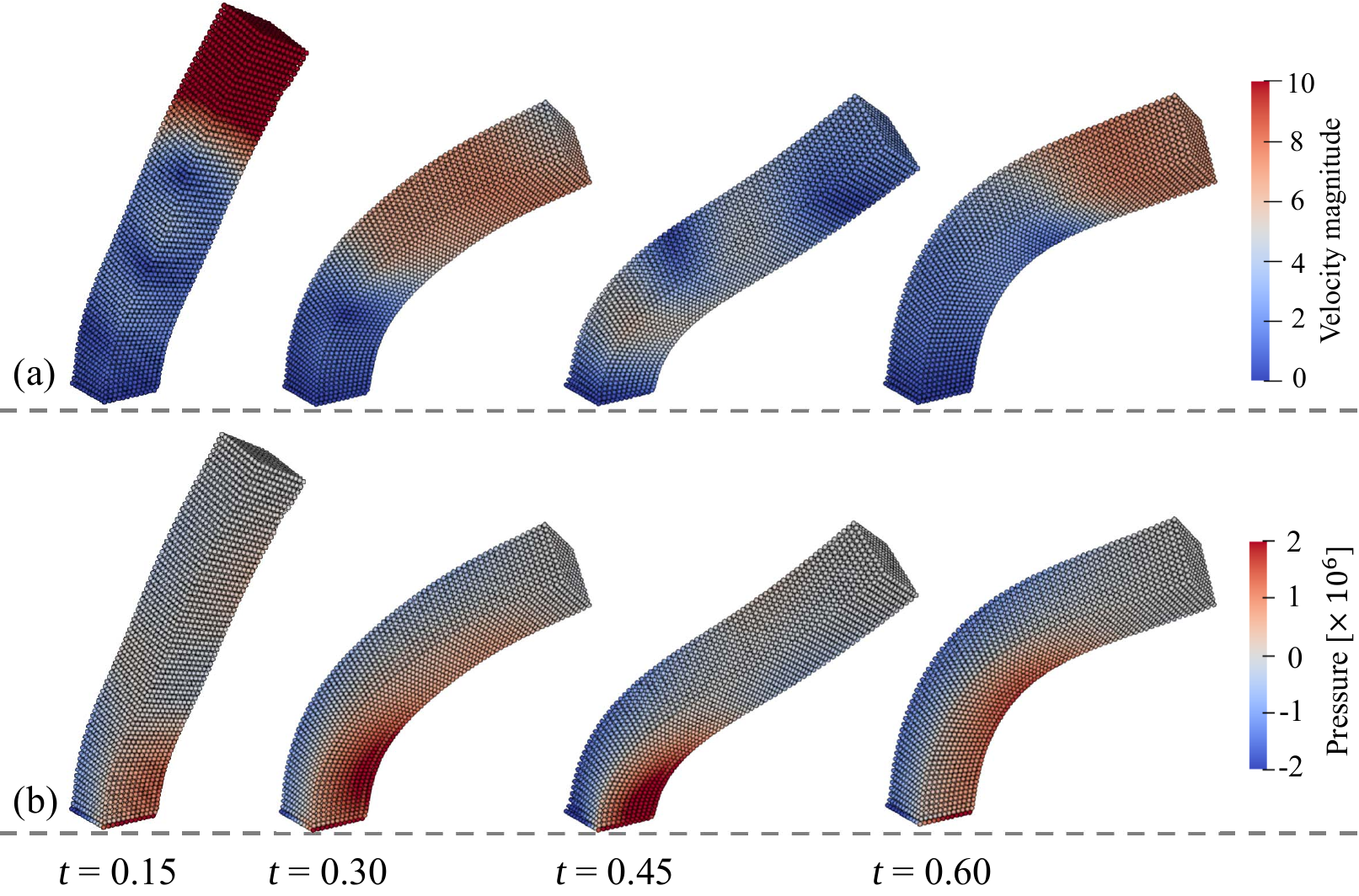}
	\caption{3D bending column: evolution of (a) velocity magnitude and (b) pressure with the present SPH-GNOG. Here, $L/dp=12$.}
	\label{figs:bending-column-pressure}
\end{figure}

Fig. \ref{figs:bending-column-stress} shows the particle configuration colored by von Mises stress over time. In this example, the results obtained from TLSPH (Fig. \ref{figs:bending-column-stress}a) \cite{wu2023essentially} are considered as a benchmark.
We also compare our results with those calculated using SPH-OG (Fig. \ref{figs:bending-column-stress}b) and SPH-ENOG (Fig. \ref{figs:bending-column-stress}c). 
Since the artificial stress formulation is defined in 2D and lacks a 3D counterpart \cite{gray2001sph}, this example does not include a comparison with SPH-OAS results.
It can be seen that the results obtained using SPH-OG exhibit severe zigzag modes and non-physical fractures. 
Although SPH-ENOG eliminates these zigzag modes and non-physical fractures, the position and shape of the column differ significantly from the TLSPH results. 
A detailed analysis reveals that in the SPH-ENOG results, once the column bends, it struggles to recover, due to poor conservation of angular momentum in SPH-ENOG, leading to significant errors in bending-dominated cases. 
In contrast, the results obtained using our proposed SPH-GNOG method (Fig. \ref{figs:bending-column-stress}d) closely match the TLSPH results in terms of both column position and stress distribution.
Fig. \ref{figs:bending-column-pressure} also presents the velocity magnitude and pressure distribution of the column at four different time points, where negative pressure values indicate tension.

\begin{figure}[htb!]
	\centering
	\includegraphics[trim = 0cm 0cm 0cm 0cm, clip,width=0.8\textwidth]{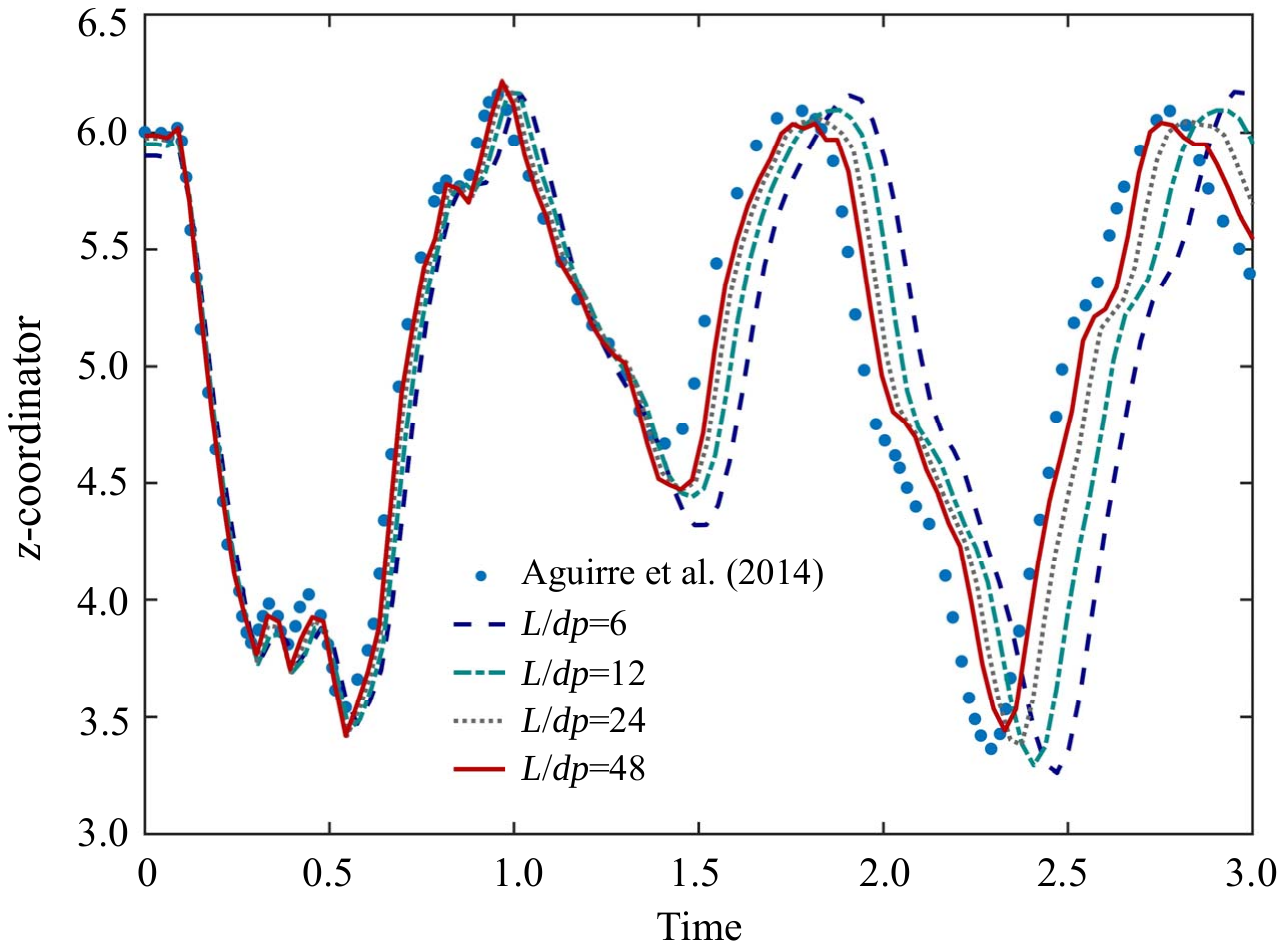}
	\caption{3D bending column: time history of the vertical position $z$ observed at node $s$ obtained by SPH-GNOG with for different spatial resolutions. The results are compared with those obtained with FVM by Aguirre et al. \cite{aguirre2014vertex}}.
	\label{figs:bending-column-position}
\end{figure}

Next, we tested the convergence and accuracy of the algorithm. As shown in Fig. \ref{figs:bending-column-position}, the variation of the z-coordinate of observation point $s$ over time is displayed with four different resolutions, namely $L/dp=6,12,24$ and 48. 
The computational results obtained by previous researchers using the Finite Volume Method (FVM) are also presented as reference values. 
It can be observed that as the resolution increases, the results of the proposed method gradually converge to those obtained by the FVM.
\subsection{2D colliding rubber rings}
\label{rubber-rings}

\begin{figure}[htb!]
	\centering
	\includegraphics[trim = 0cm 0cm 0cm 0cm, clip,width=0.7\textwidth]{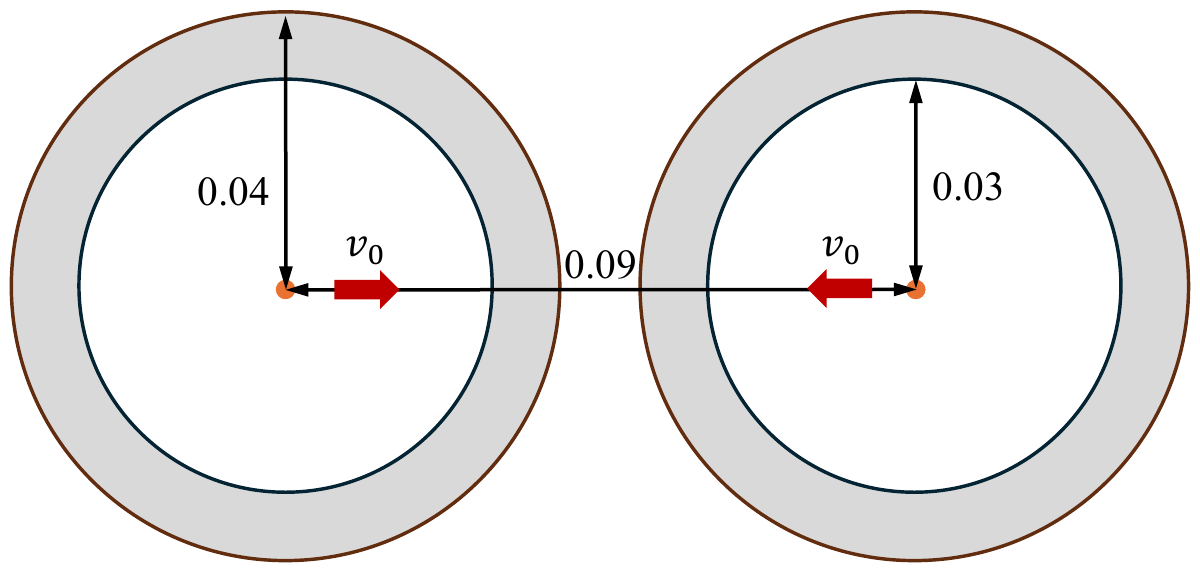}
	\caption{2D colliding rubber rings: model setup.}
	\label{figs:2D-ring-setup}
\end{figure}

\begin{figure}[htb!]
	\centering
	\includegraphics[trim = 0cm 0cm 0cm 0cm, clip,width=0.5\textwidth]{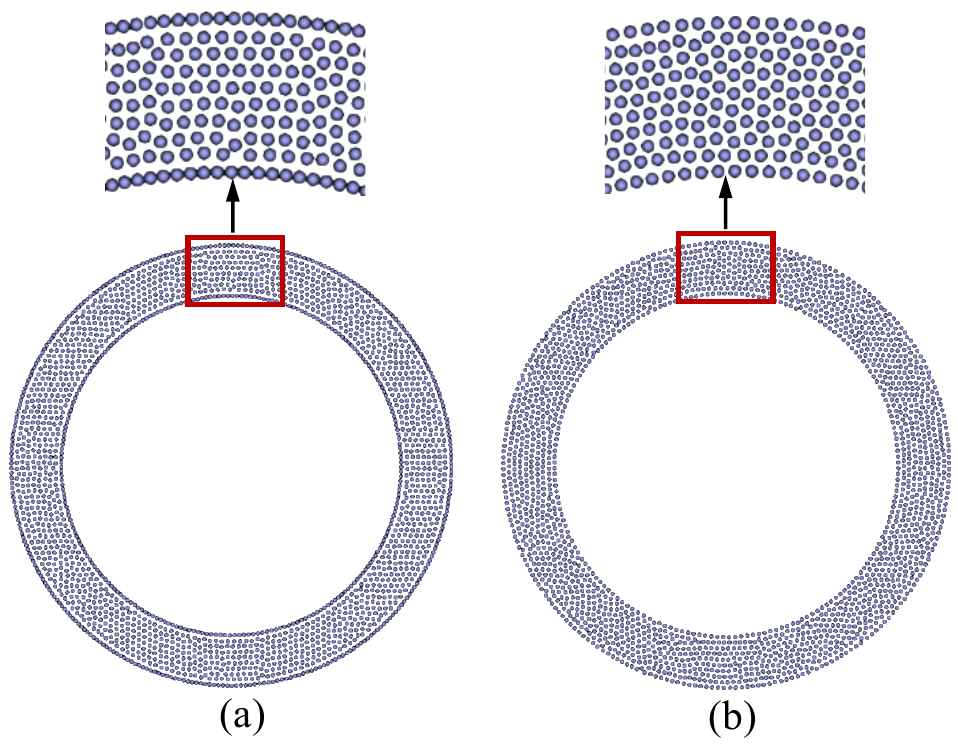}
	\caption{2D colliding rubber rings: initial particle distribution in the study of Zhang et al. \cite{zhang2024essentially} and (b) this study.}
	\label{figs:2D-ring-level-set}
\end{figure}

Referring to \cite{gray2001sph, monaghan2000sph, zhang2017generalized}, this section simulates the collision of two rings. When the two rings collide, significant tensile forces are generated. 
This section demonstrates that the present method can eliminate numerical instabilities, namely zigzag modes and non-physical fractures, during the simulation process. As shown in Fig. \ref{figs:2D-ring-setup}, two rings with an inner radius of 0.03 and an outer radius of 0.04 move towards each other with an initial velocity $v_0$ (the relative velocity of the two rings is 2$v_0$), and the initial distance between the centers of the two rings is 0.09.
According to the Zhang et al. \cite{zhang2024essentially}, we employ an irregular initial particle distribution to address a generic scenario. The choice of an irregular distribution is made to avoid the simplification inherent in a radial particle distribution, especially when dealing with complex geometries \cite{zhang2017generalized}. 
However, unlike the literature \cite{zhang2024essentially}, this study adopts a level-set correction scheme \cite{yu2023level} to achieve a uniform particle distribution at the model boundaries, as shown in Fig. \ref{figs:2D-ring-level-set}.
The material parameters are specified as follows: density $\rho_0=1200$, Young's modulus $E=1\times 10^7$, and Poisson's ratio $\nu = 0.4$. The initial particle spacing is set to 0.001.

\begin{figure}[htb!]
	\centering
	\includegraphics[trim = 0cm 0cm 0cm 0cm, clip,width=1.0\textwidth]{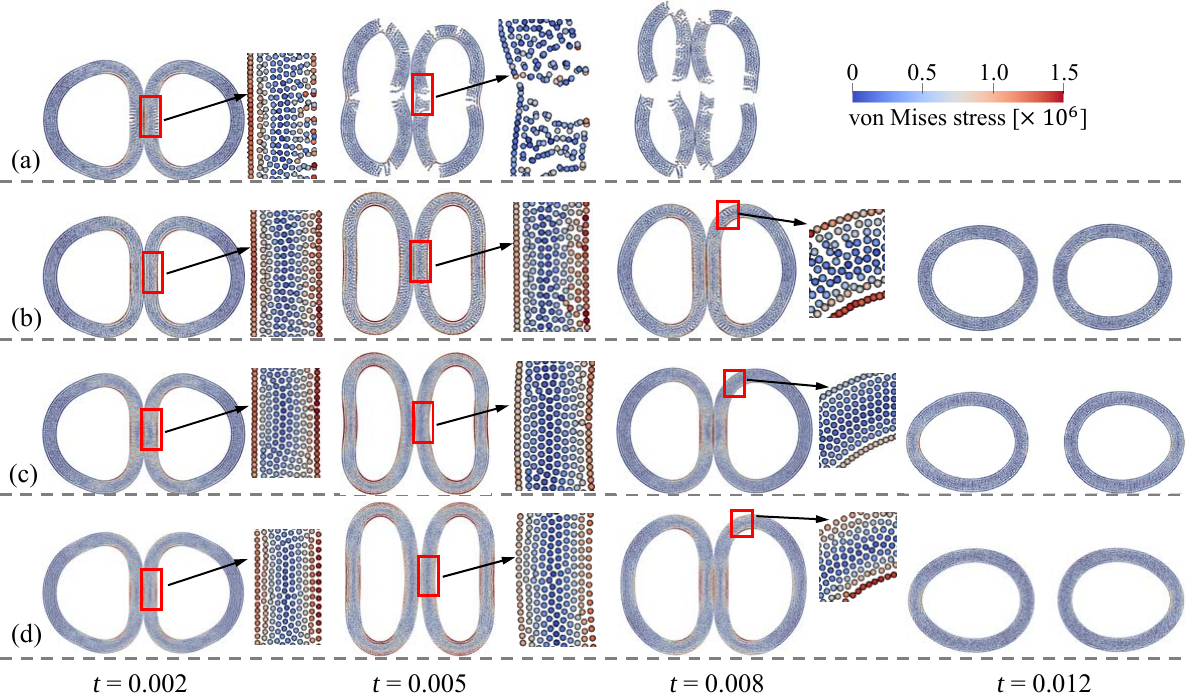}
	\caption{2D colliding rubber rings: evolution of particle configuration with time ($t=0.002$, 0.005, 0.008 and 0.012). The results are obtained by four different SPH strategies, i.e., (a) SPH-OG \cite{zhang2024essentially}, (b) SPH-OAS \cite{gray2001sph, zhang2024essentially}, (c) SPH-ENOG \cite{zhang2024essentially}, and (d) SPH-GNOG. The initial velocity magnitude $v_0=0.06c_0$ and the particles are colored by von Mises stress.}
	\label{figs:2D-ring-v12}
\end{figure}

Figure \ref{figs:2D-ring-v12} illustrates the evolution of particle configurations for SPH-OG, SPH-OAS, SPH-ENOG, and the proposed SPH-GNOG at an initial velocity magnitude of $v_0=0.06c_0$. SPH-OG clearly exhibits severe non-physical fractures and zigzag patterns early in the computation ($t=0.002$), causing the calculation to nearly halt. In contrast, SPH-OAS manages to suppress these fractures, maintaining a uniform particle distribution initially ($t=0.002$). However, over time, the particle configuration and von Mises stress begin to display a zigzag pattern. The proposed SPH-GNOG, similar to SPH-ENOG, ensures a uniform particle and stress distribution throughout the entire computation, completely eliminating numerical instabilities.

\begin{figure}[htb!]
	\centering
	\includegraphics[trim = 0cm 0cm 0cm 0cm, clip,width=1.0\textwidth]{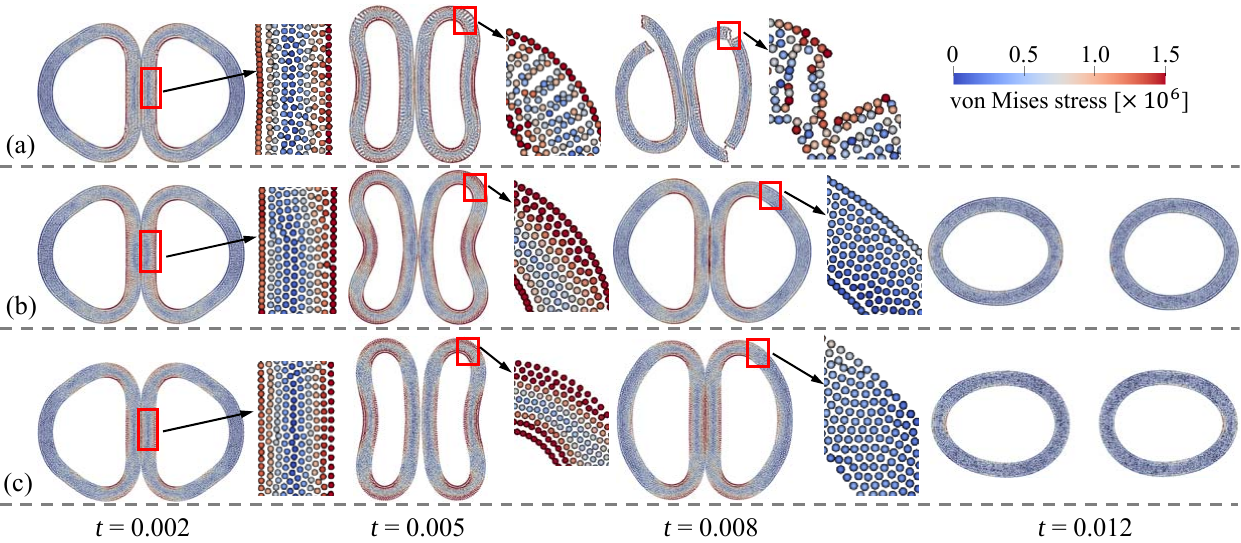}
	\caption{2D colliding rubber rings: evolution of particle configuration with time ($t=0.002$, 0.005, 0.008 and 0.012). The results are obtained by three different SPH strategies, i.e., (a) SPH-OAS \cite{zhang2024essentially}, (b) SPH-ENOG \cite{gray2001sph, zhang2024essentially}, and (c) SPH-GNOG. The initial velocity magnitude $v_0=0.08c_0$ and the particles are colored by von Mises stress.}
	\label{figs:2D-ring-v16}
\end{figure}

Furthermore, we increased the initial velocity to $v_0=0.08c_0$ to further test the stability of the algorithm. As shown in Fig. \ref{figs:2D-ring-v16}, for SPH-OAS, not only zigzag patterns but also numerical fractures appear when $t \geqslant 0.005$. This is consistent with the literatures \cite{zhang2017generalized, lobovsky2007smoothed, zhang2024essentially}, which state that the SPH-OAS method fails when the material deformation is too large or when the material has a high Poisson's ratio. Same to SPH-ENOG, the proposed SPH-GNOG performs well even at such a high initial velocity, with all numerical instabilities being perfectly eliminated. The shape of the ring is highly consistent with the results from SPH-ENOG, demonstrating the stability, robustness, and accuracy of the proposed SPH-GNOG.

\begin{figure}[htb!]
	\centering
	\includegraphics[trim = 0cm 0cm 0cm 0cm, clip,width=0.75\textwidth]{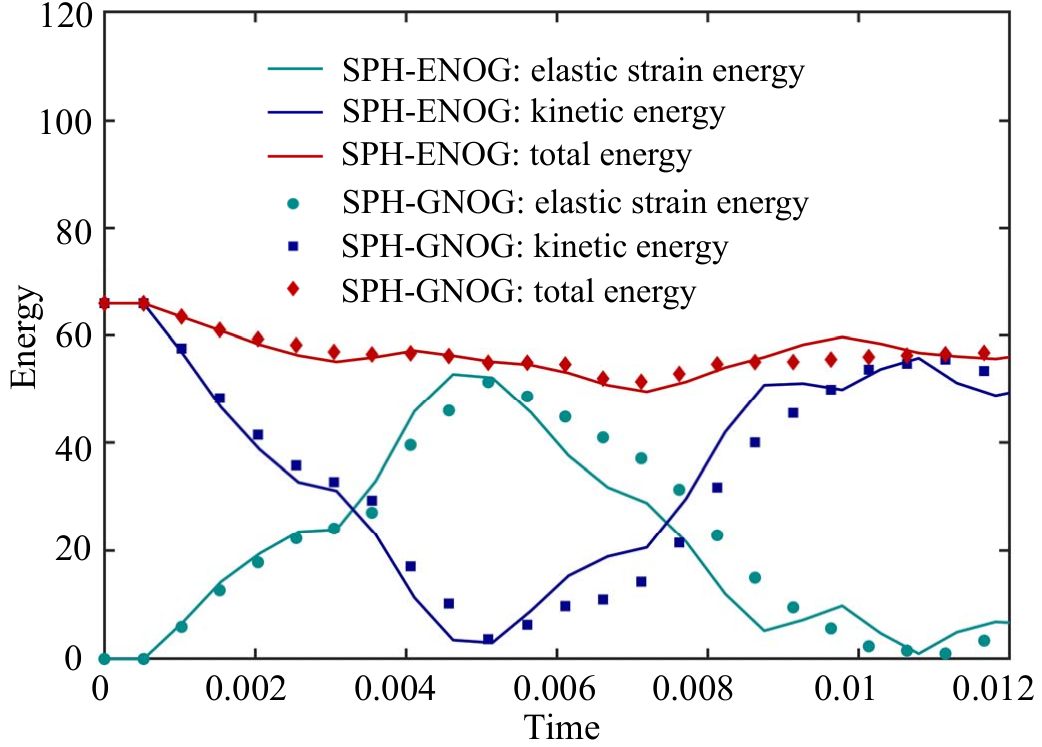}
	\caption{2D colliding rubber rings: the variation of elastic strain energy, kinetic energy, and total energy over time for the left ring with the SPH-ENOG \cite{zhang2024essentially} (represented by solid lines) and SPH-GNOG (represented by dots). The initial velocity magnitude $v_0=0.06c_0$.}
	\label{figs:2D-ring-energy}
\end{figure}

Figure \ref{figs:2D-ring-energy} illustrates the energy variation of the rings. Specifically, it shows the changes in elastic strain energy, kinetic energy, and total energy over time for the left ring calculated using SPH-ENOG (represented by solid lines) and SPH-GNOG (represented by dots).
The energy of the right ring is identical to that of the left ring.
It can be seen that the overall variation trend of the SPH-GNOG results is consistent with that of SPH-ENOG. The kinetic energy decreases its minimum around $t=0.005$ and then gradually increases, while the elastic strain energy follows an opposite pattern. Although there are some differences in the specific values of kinetic energy and elastic strain energy between the two methods, due to different initial particle distributions and formulations used, the total energy is very close. For SPH-ENOG, the total energy decreases from an initial value of 65.88 to 55.61 by the end of the simulation, while for SPH-GNOG, the total energy decreases from an initial value of 65.88 to 56.74.

\subsection{3D colliding rubber balls}
\label{rubber-balls}
The 2D colliding rubber rings are extended to 3D to test the proposed SPH-GNOG in 3D scenarios. In this setup, two hollow rubber balls with an inner radius of 0.03 and an outer radius of 0.04 move toward each other. They start with an initial center-to-center distance of 0.09 and an initial velocity magnitude of $v_0$. The particles are spaced at $dp=0.001$, and a uniform particle distribution is achieved using a level-set based pre-processing technique \cite{yu2023level}. Material parameters are chosen according to Section \ref{rubber-rings}.

\begin{figure}[htb!]
	\centering
	\includegraphics[trim = 0cm 0cm 0cm 0cm, clip,width=0.95\textwidth]{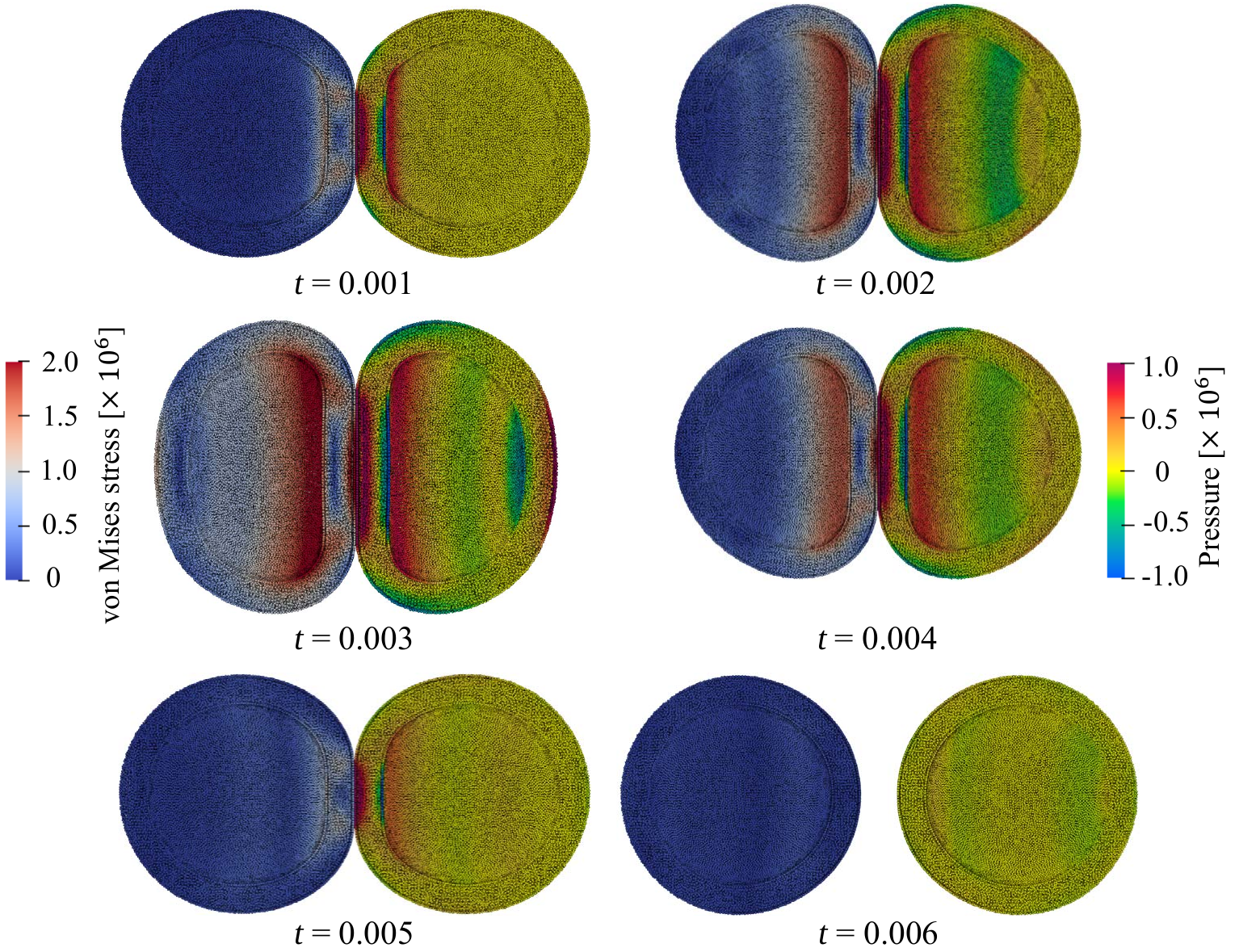}
	\caption{3D colliding rubber balls: evolution of particle configuration with time. The results are obtained by the present SPH-GNOG. The initial velocity magnitude $v_0=0.08c_0$. The left ball is colored by von Mises stress, and the right ball is colored by pressure.}
	\label{figs:3D-ball-stress}
\end{figure}

Fig. \ref{figs:3D-ball-stress} presents the collision process of two balls at various time points ($t=0.001$, 0.002, 0.003, 0.04, 0.005, and 0.006) using the proposed SPH-ENOG. The initial velocity is set to $v_0=0.08c_0$. For better visualization, only half of each ball is shown separately. 
The left ball is colored according to von Mises stress, while the right ball is colored according to pressure, with negative values indicating tension.
The von Mises stress and pressure distributions are clearly smooth, and the particle configuration remains uniform, indicating that numerical fractures and zigzag patterns are entirely eliminated.

\begin{figure}[htb!]
	\centering
	\includegraphics[trim = 0cm 0cm 0cm 0cm, clip,width=0.7\textwidth]{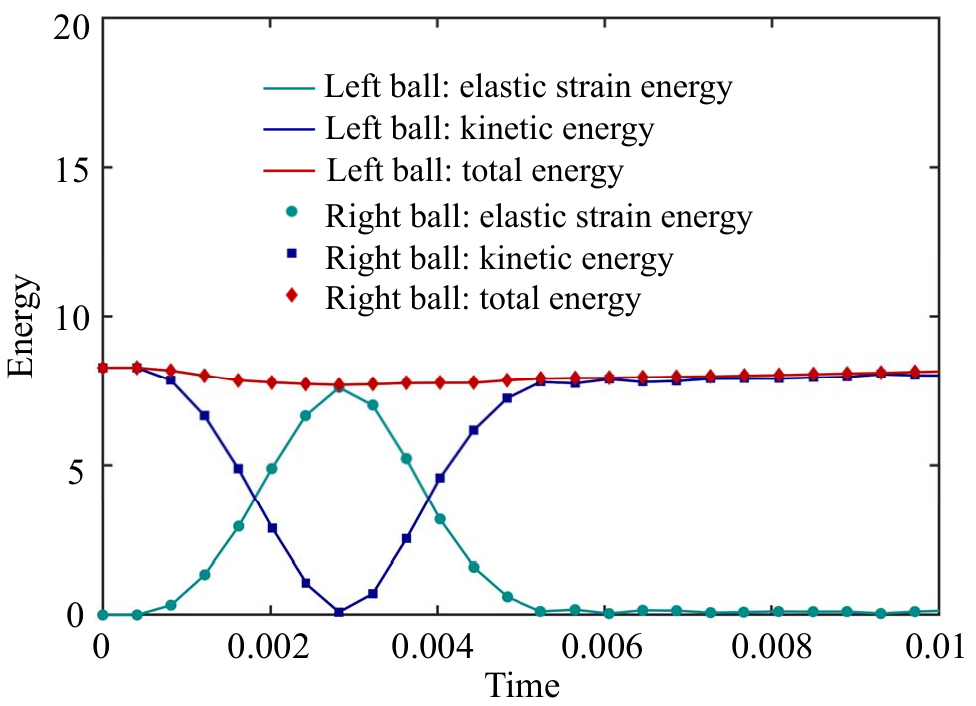}
	\caption{3D colliding rubber balls: the variation of elastic strain energy, kinetic energy, and total energy over time for the left (represented by solid lines) and right (represented by dots) balls. The initial velocity magnitude $v_0=0.08c_0$.}
	\label{figs:3D-ball-energy}
\end{figure}

Fig. \ref{figs:3D-ball-energy} shows the variation of elastic strain energy, kinetic energy, and total energy over time for the left and right balls. It can be seen that kinetic energy gradually decreases after the balls make contact, reaching its minimum value at $t \approx 0.003$, while the elastic strain energy peaks at this time. 
Subsequently, the release of elastic strain energy causes the kinetic energy to gradually increase and remain relatively constant after the balls separate. 
Due to numerical dissipation, there is a slight loss in total energy during the simulation, decreasing from the initial value of 8.27 to 8.16 by the end of the simulation.
\subsection{Spinning plate}
\label{spinning-plate}
\begin{figure}[htb!]
	\centering
	\includegraphics[trim = 0cm 0cm 0cm 0cm, clip,width=0.35\textwidth]{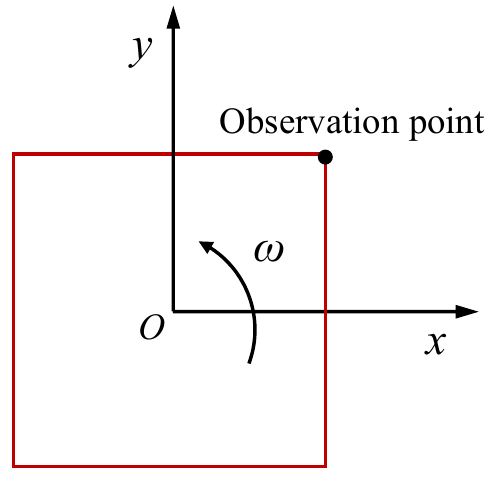}
	\caption{Spinning plate: model setup.}
	\label{figs:spinning-plate-setup}
\end{figure}

We further test the effectiveness of our algorithm in eliminating numerical instabilities through a case entirely dominated by tensile forces.
Fig. \ref{figs:spinning-plate-setup} shows the configuration for the spinning plate. 
A square plate with a side length of one rotates around its center without any initial deformation or constraints, with an angular velocity of $\omega=50$. The material parameters are set according to \cite{lee2013development}: density $\rho_0=1100$, Young's modulus $E=1.7 \times 10^{7}$, and Poisson's ratio $\nu=0.45$.
A monitoring point is positioned at the top-right corner of the plate to record displacement and velocity.
The initial particle spacing is set to 0.05.

\begin{figure}[htb!]
	\centering
	\includegraphics[trim = 0cm 0cm 0cm 0cm, clip,width=0.9\textwidth]{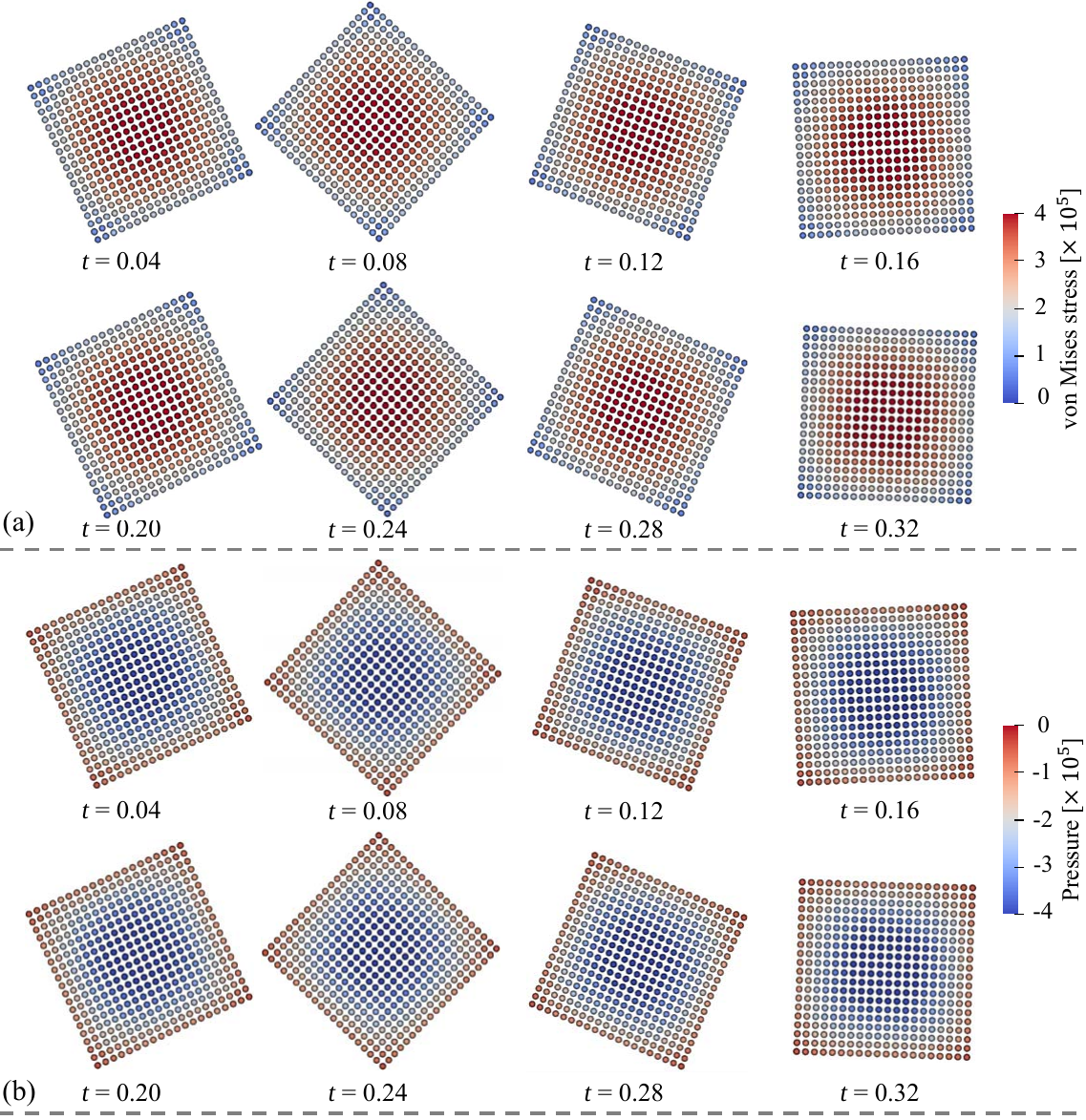}
	\caption{Spinning plate: the evolution of particle configuration at different instants. The particles are colored by (a) von Mises stress and (b) pressure respectively.}
	\label{figs:spinning-plate-stress}
\end{figure}

Fig. \ref{figs:spinning-plate-stress} shows the variation in the distribution of von Mises stress and pressure over time. It can be seen that not only is the particle distribution very uniform, but the stress and pressure distributions are also very smooth. This indicates that our method can completely eliminate zigzag modes and non-physical fractures even in scenarios dominated by tensile forces.

\begin{figure}[htb!]
	\centering
	\includegraphics[trim = 0cm 0cm 0cm 0cm, clip,width=0.95\textwidth]{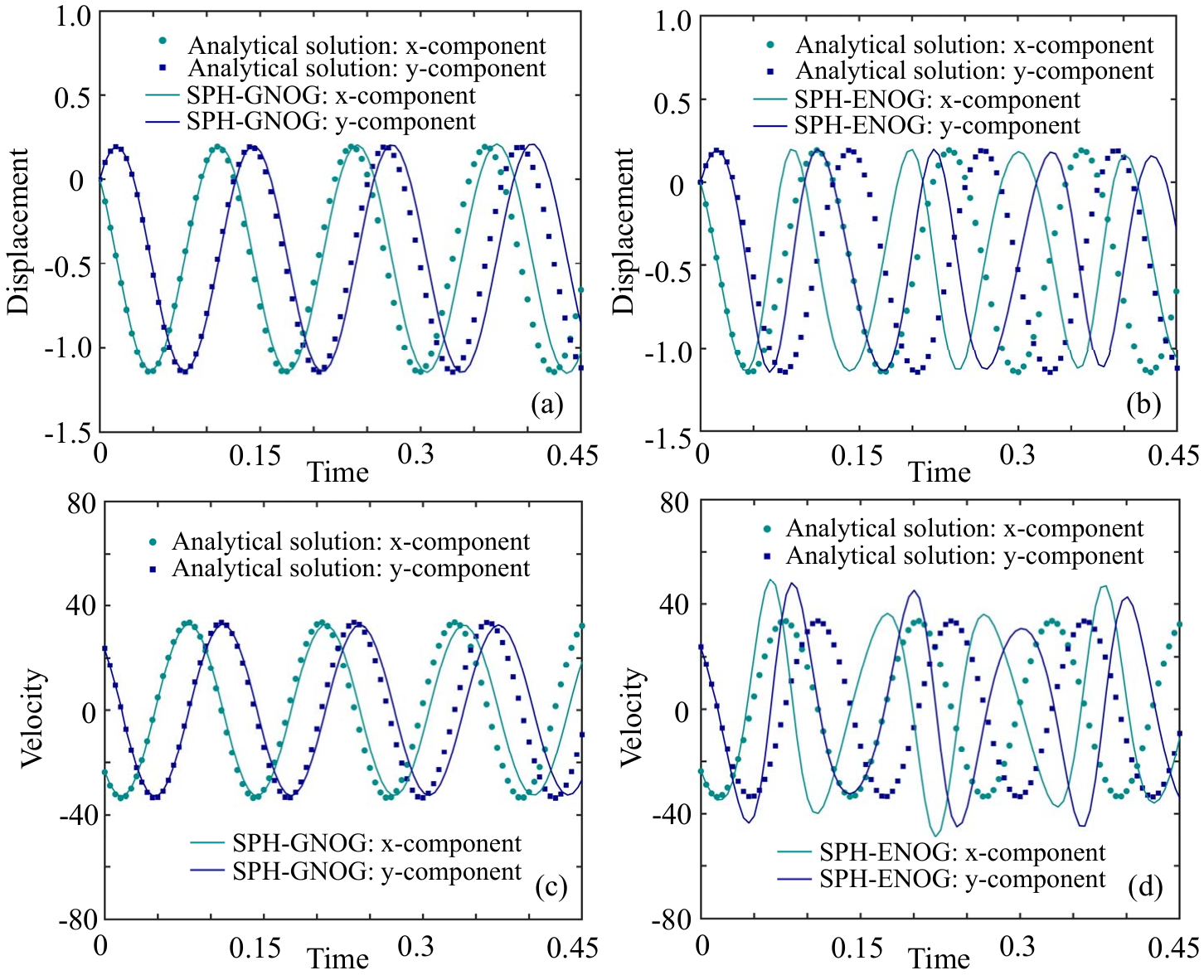}
	\caption{Spinning plate: the temporal evolution of displacement and velocity for the observation point with SPH-GNOG and SPH-ENOG. (a) The $x$ and $y$ components of displacement with SPH-GNOG; (b) the $x$ and $y$ components of displacement with SPH-ENOG \cite{zhang2024essentially}; (c) the $x$ and $y$ components of velocity with SPH-GNOG; (d) the $x$ and $y$ components of velocity with SPH-ENOG \cite{zhang2024essentially}.}
	\label{figs:spinning-plate-disp}
\end{figure}

\begin{figure}[htb!]
	\centering
	\includegraphics[trim = 0cm 0cm 0cm 0cm, clip,width=0.95\textwidth]{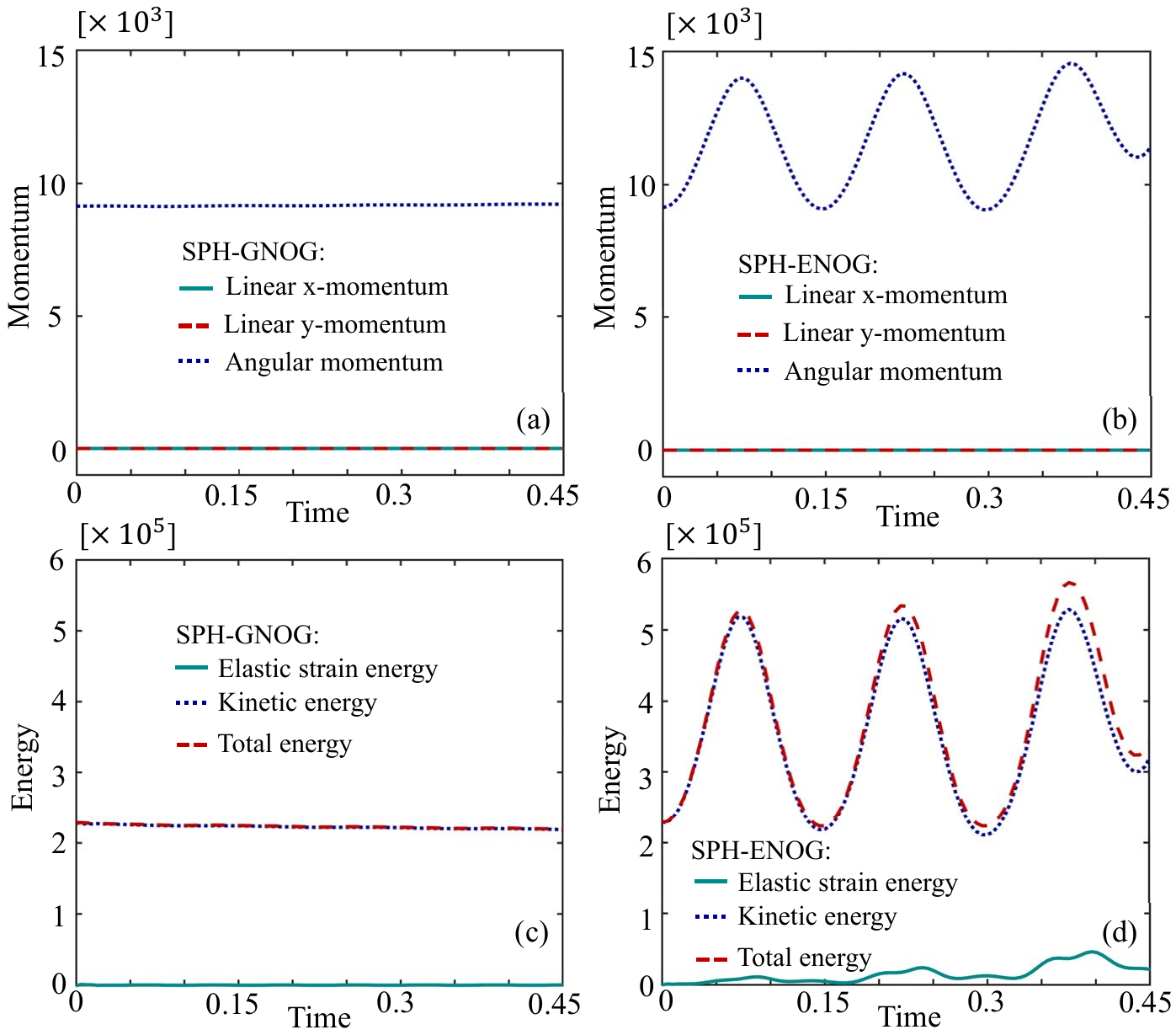}
	\caption{Spinning plate: the temporal evolution of the system's momentum and energy SPH-GNOG and SPH-ENOG. (a) The total linear momentum and angular momentum with SPH-GNOG; (b) the total linear momentum and angular momentum with SPH-ENOG \cite{zhang2024essentially}; (c)  the elastic strain energy, kinetic energy and total energy with SPH-GNOG; (d) the elastic strain energy, kinetic energy and total energy with SPH-ENOG \cite{zhang2024essentially}.}
	\label{figs:spinning-plate-energy}
\end{figure}

Fig. \ref{figs:spinning-plate-disp} compares the displacement and velocity variations at the observation point with analytical values. 
It is evident that the results obtained using the SPH-GNOG method show good agreement with the theoretical values in both displacement (Fig. \ref{figs:spinning-plate-disp}a) and velocity (Fig. \ref{figs:spinning-plate-disp}c) over time. In contrast, the SPH-ENOG method exhibits significant deviations from the theoretical values \cite{zhang2024essentially}.
As observed from Fig. \ref{figs:spinning-plate-energy}a, our method effectively conserves linear momentum. 
While angular momentum is not strictly conserved, the level of conservation remains quite satisfactory.
On the other hand, although the SPH-ENOG method can eliminate numerical instabilities, it poorly conserves angular momentum (Fig. \ref{figs:spinning-plate-energy}b) \cite{zhang2024essentially}. 
This poor conservation is the primary reason for the large discrepancies in displacement and velocity compared to the theoretical values.
Fig. \ref{figs:spinning-plate-energy}c shows the variation of elastic strain energy, kinetic energy, and total energy over time using the SPH-GNOG. As time progresses, the total energy slightly decreases as the kinetic energy decreases due to numerical dissipation, while the the elastic strain energy remains close to zero. 
Conversely, the SPH-ENOG exhibits significant energy conservation issues (Fig. \ref{figs:spinning-plate-energy}d).
These results indicate that in scenarios dominated by angular momentum, the performance of the SPH-GNOG is significantly superior to that of the SPH-ENOG.

\subsection{Round Taylor bar}
\label{round-taylor-bar}
\begin{figure}[htb!]
	\centering
	\includegraphics[trim = 0cm 0cm 0cm 0cm, clip,width=0.8\textwidth]{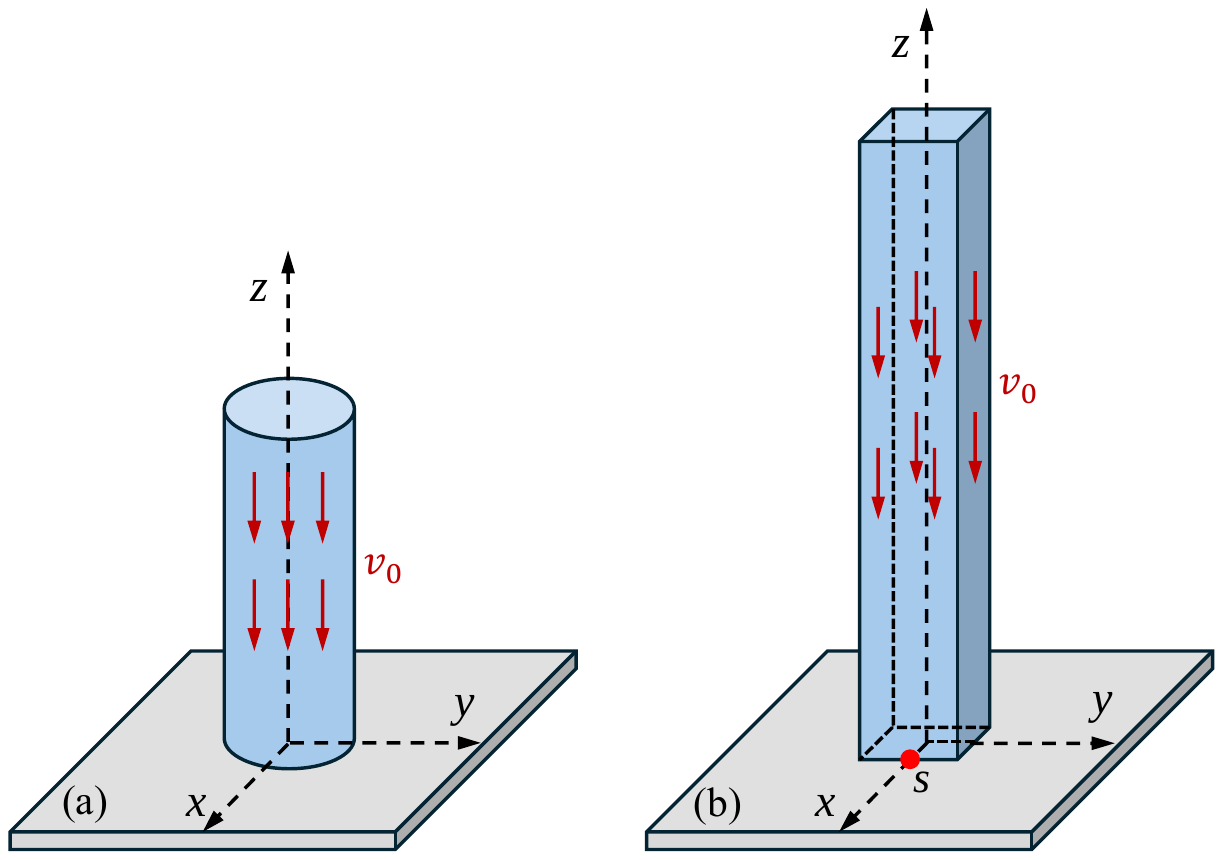}
	\caption{Model setup for (a) round Taylor bar and (b) square Taylor bar.}
	\label{figs:taylor-bar-setup}
\end{figure}

A classic impact problem, i.e., an aluminum bar impact on a rigid wall introduced by Taylor \cite{taylor1948use} to measure yield properties, is widely used to validate the effectiveness of the proposed model for elastoplastic materials \cite{chen1996reproducing, lee2014development}.
As shown in Fig. \ref{figs:taylor-bar-setup}a, a cylindrical bar, with initial radius $R=3.91 \times 10^{-3}$ and length $2.346 \times 10^{-2}$, impact against a rigid frictionless wall with velocity $\mathbf v_0=(0,0,-373)$.
The moment at $t=0$ is when the bar is just about to come into contact with the wall.
A $J_2$ plastic model with perfect plasticity is used to describe the material response \cite{chen1996reproducing}.
The material parameters are \cite{chen1996reproducing}: density $\rho_0=2700$, Young's modulus $E=7.82 \times 10^{10}$, Poisson's ratio $\nu=0.3$, and yield stress $\sigma_Y=2.9 \times 10^{8}$.

\begin{figure}[htb!]
	\centering
	\includegraphics[trim = 0cm 0cm 0cm 0cm, clip,width=1.0\textwidth]{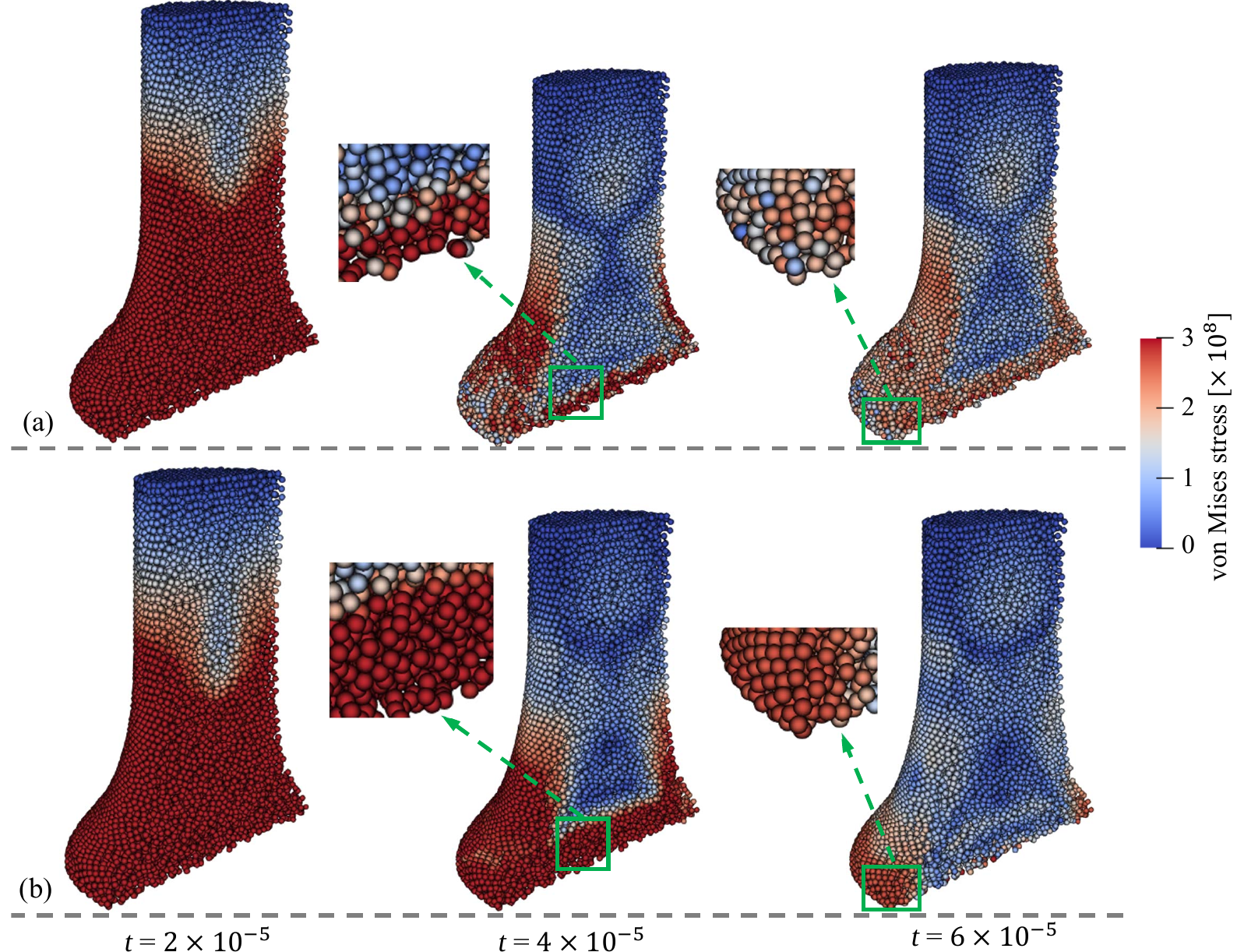}
	\caption{Round Taylor bar: evolution of particle configuration with time. The results are obtained by (a) SPH-OG and (b) SPH-GNOG. The initial particle spacing $dp=R/12$ and the particles are colored by von Mises stress.}
	\label{figs:round-bar-stress}
\end{figure}

\begin{figure}[htb!]
	\centering
	\includegraphics[trim = 0cm 0cm 0cm 0cm, clip,width=1\textwidth]{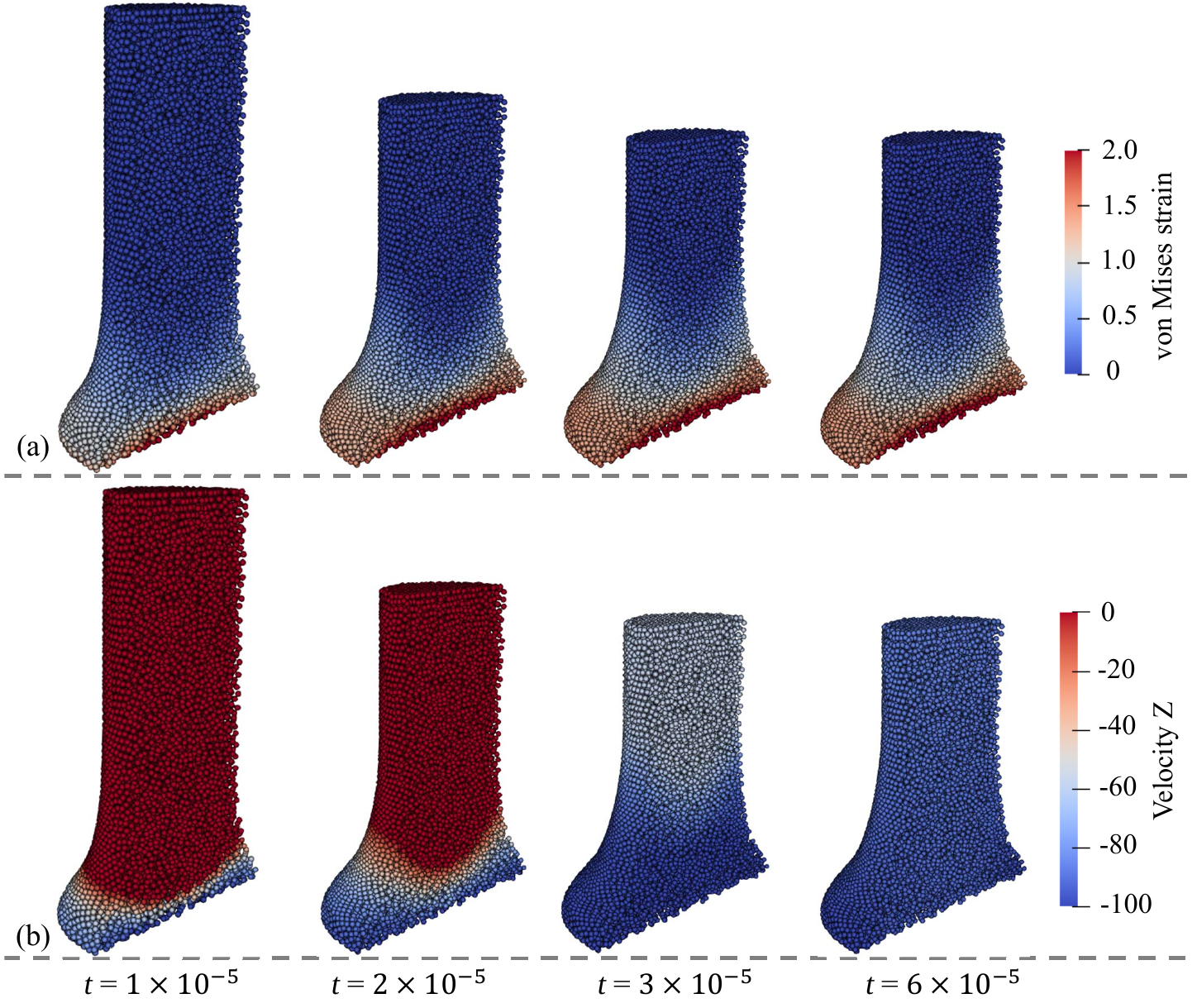}
	\caption{Round Taylor bar: evolution of (a) von Mises strain and (b) z-component of velocity  with the present SPH-GNOG. Here, $dp=R/12$.}
	\label{figs:round-bar-strain}
\end{figure}

Fig. \ref{figs:round-bar-stress} illustrates the particle configuration (colored by von Mises stress) during the high-speed impact of a cylindrical bar, simulated using SPH-OG (Fig. \ref{figs:round-bar-stress}a) and SPH-GNOG (Fig. \ref{figs:round-bar-stress}b). Given the cylindrical shape of the initial model, the particles are isotropically distributed on the horizontal cross-section. As a result, no zigzag pattern in particle distribution is observed, even with SPH-OG. 
However, a zigzag phenomenon in stress distribution is evident in the SPH-OG results, while the proposed SPH-GNOG achieves a smoother stress distribution. 
This highlights the effectiveness of the proposed algorithm in addressing hourglass modes in plastic deformation problems.
Fig. \ref{figs:round-bar-strain} presents snapshots of the von Mises strain and z-direction velocity distribution at different time points during the simulation using the present SPH-GNOG.

\begin{figure}[htb!]
	\centering
	\includegraphics[trim = 0cm 0cm 0cm 0cm, clip,width=1\textwidth]{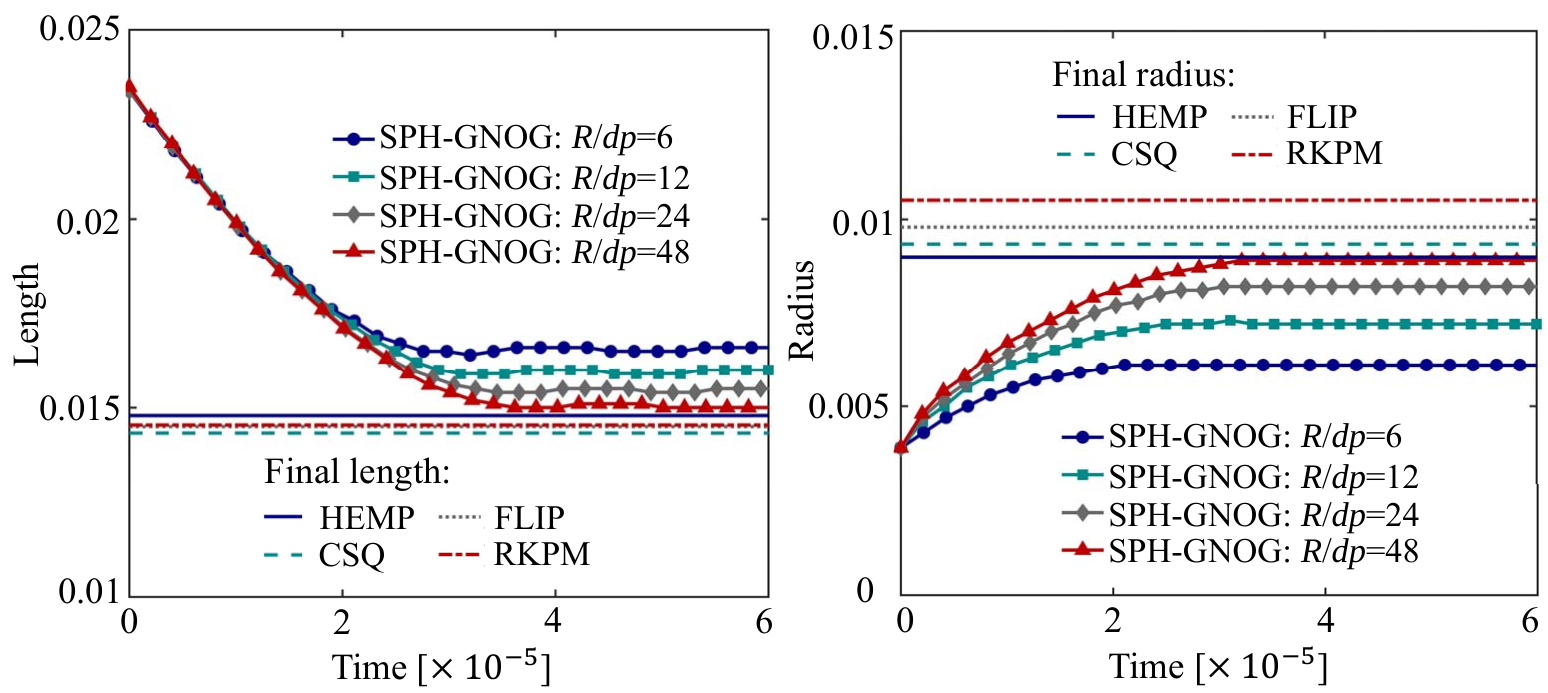}
	\caption{Round Taylor bar: temporal evolution of length and radius of the bar at various resolutions. The results are compared with the final values of length and radius obtained from HEMP \cite{wilkins1973impact}, CSQ \cite{predebon1991inclusion}, FLIP \cite{sulsky1995application}, and RKPM \cite{chen1996reproducing}.}
	\label{figs:round-bar-radius}
\end{figure}

Fig. \ref{figs:round-bar-radius} shows the changes in the length and radius of the bar over time at different resolutions, represented by dotted lines. 
The lines without dots represent the final length and radius of the bar obtained by other numerical methods, specifically finite element predictions using HEMP \cite{wilkins1973impact}, finite difference results by CSQ \cite{predebon1991inclusion}, particle in cell solution with FLIP \cite{sulsky1995application}, and results obtained by reproducing kernel particle methods (RKPM) \cite{chen1996reproducing}.
Clearly, as the resolution increases, the final length and radius calculated by the present SPH-GNOG gradually converge to the results in literatures.
\subsection{Square Taylor bar}
\label{square-taylor-bar}
Next, we investigate the impact of a square cross-section Taylor bar with a rigid wall.
As shown in Fig. \ref{figs:taylor-bar-setup}b, the square bar has an initial height of $H=0.03$, a square cross-section with a side length of $L=0.006$, and an initial velocity of $\mathbf v_0=(0,0,-227)$ \cite{haider2017first}. 
An observation point $s$ is set at the coordinate $(0.003,0,0)$ to measure the displacement.
The $J_2$ plasticity model with a linear hardening law is adopted \cite{haider2017first}. 
The material parameters are as follows: density $\rho_0=8930$, Young's modulus $E=1.17 \times 10^{11}$, Poisson's ratio $\nu=0.35$, yield stress $\sigma_Y=4 \times 10^{8}$, and hardening modulus $\kappa =1 \times 10^{8}$.

\begin{figure}[htb!]
	\centering
	\includegraphics[trim = 0cm 0cm 0cm 0cm, clip,width=1.0\textwidth]{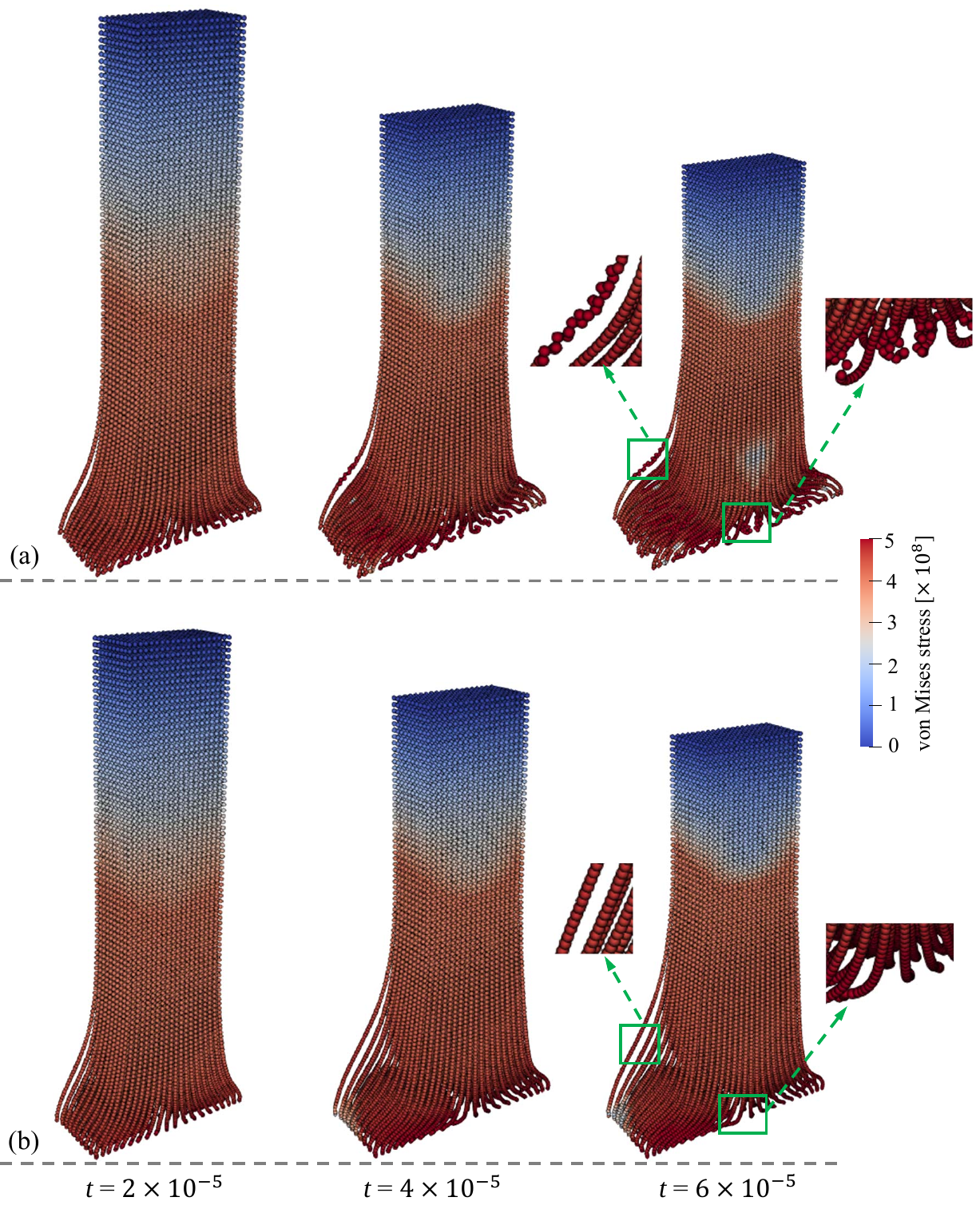}
	\caption{Square Taylor bar: evolution of particle configuration with time. The results are obtained by (a) SPH-OG and (b) the present SPH-GNOG. The initial particle spacing $dp=W/20$ and the particles are colored by von Mises stress.}
	\label{figs:square-bar-stress}
\end{figure}

\begin{figure}[htb!]
	\centering
	\includegraphics[trim = 0cm 0cm 0cm 0cm, clip,width=1\textwidth]{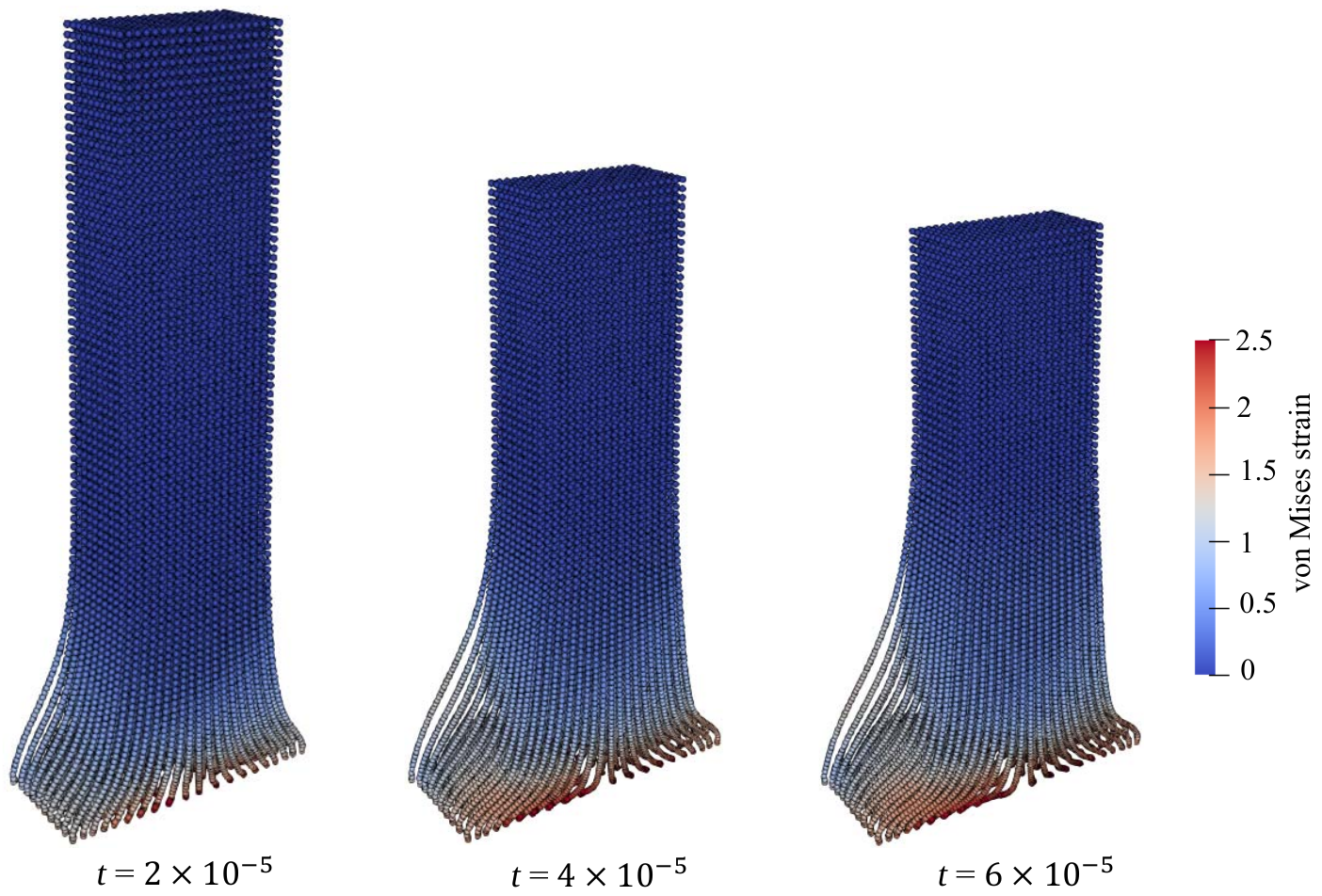}
	\caption{Square Taylor bar: evolution of von Mises strain with the present SPH-GNOG. Here, $dp=W/20$.}
	\label{figs:square-bar-strain}
\end{figure}

Fig. \ref{figs:square-bar-stress} shows the evolution of the particle configuration of the square bar over time when using SPH-OG and SPH-GNOG, colored by von Mises stress. 
With SPH-OG, noticeable particle disorder is observed, whereas SPH-GNOG eliminates these issues, demonstrating a significant advantage over traditional methods.
Fig. \ref{figs:square-bar-strain} presents snapshots of the von Mises strain at different time points during the simulation using the present SPH-GNOG.

The final x-coordinate of the observation point $s$ is recorded at different particle resolutions and compared with the results obtained by Haider et al. \cite{haider2017first} using the upwind cell-centered Total Lagrangian scheme. From Table \ref{square-bar-x}, it can be seen that as the resolution increases, the x-coordinate of point $s$ rapidly converges to the results of Haider et al \cite{haider2017first}.
\begin{table}
	\scriptsize
	\centering
	\caption{Square Taylor bar: The x-coordinate of the observation point $s$ at the final moment ($t=6 \times 10^{-5}$).}
	\begin{tabularx}{9.5cm}{@{\extracolsep{\fill}}lcccc}
		\hline
		$ \ $ & $L/dp=10$ & $L/dp=20$ & $L/dp=30$ & Haider et al. \cite{haider2017first}\\
		\hline
        x-coordinate & $4.73 \times 10^{-3}$ & $6.34 \times 10^{-3}$ & $6.87 \times 10^{-3}$ & $6.93 \times 10^{-3}$  \\
		\hline
	\end{tabularx}
	\label{square-bar-x}
\end{table}

\subsection{High-velocity impact}
\label{HVI}
\begin{figure}[htb!]
	\centering
	\includegraphics[trim = 0cm 0cm 0cm 0cm, clip,width=0.25\textwidth]{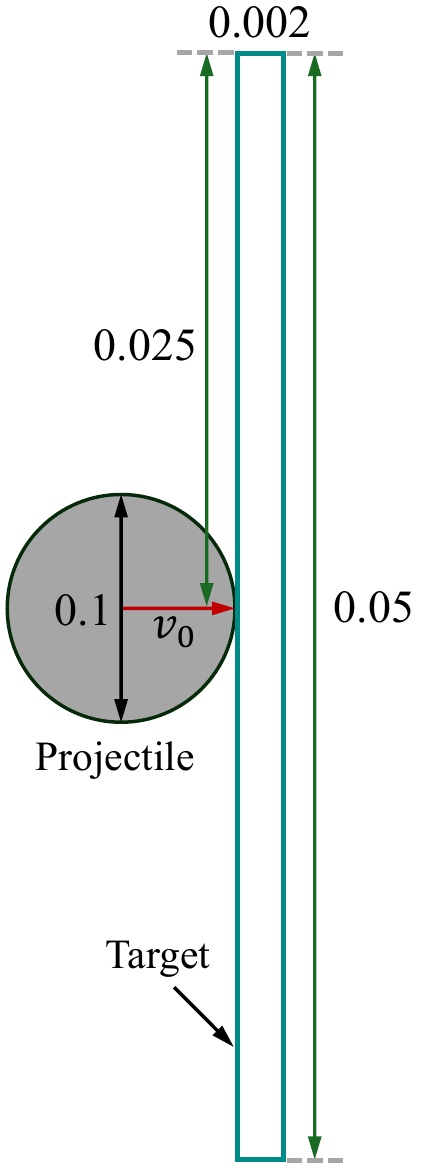}
	\caption{High-velocity impact: model setup.}
	\label{figs:HVI-setup}
\end{figure}

A challenging case, high-velocity impact problem involving a circular projectile impacting on a rectangular target \cite{howell2002free,mehra2006high}, is investigated in this section.
The simulation presented here aims to assess the effectiveness of the proposed algorithm in handling discontinuous deformation issues, particularly in scenarios involving material fragmentation.
As illustrated in Fig. \ref{figs:HVI-setup}, a circular projectile, with a diameter of 0.01 and an initial velocity of $v_0 = 3100$, impacted a rectangular target measuring 0.05 in height and 0.002 in width \cite{mehra2006high}.
Both the projectile and the target are modeled as aluminum \cite{mehra2006high}. The physical parameters are set according to previous studies \cite{mehra2006high, young2021adaptive}, i.e., 
density $\rho_0=2785$, Young's modulus $E=7.417 \times 10^{10}$, Poisson's ratio $\nu=0.344$, and sound speed $c_0 =5328$. Both the projectile and the target are modeled as elastic-perfectly plastic materials with a yield stress of $\sigma_Y = 3 \times 10^{8}$.
The initial particle spacing $dp=0.0002$.

\begin{figure}[htb!]
	\centering
	\includegraphics[trim = 0cm 0cm 0cm 0cm, clip,width=1\textwidth]{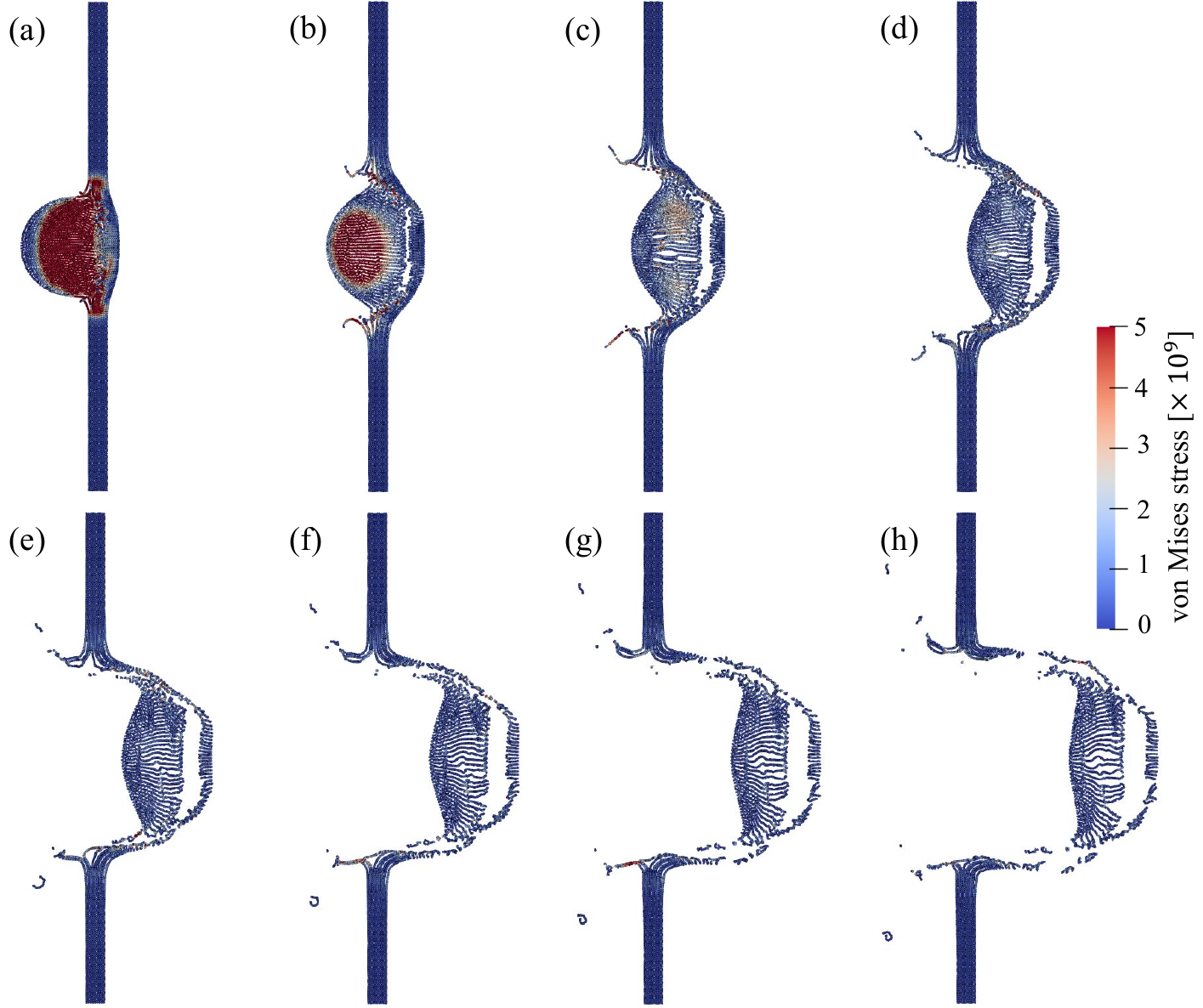}
	\caption{High-velocity impact: evolution of von Mises stress with the present SPH-GNOG at (a) $t=1 \times 10^{-6}$, (b) $t=2 \times 10^{-6}$, (c) $t=3 \times 10^{-6}$, (d) $t=4 \times 10^{-6}$, (e) $t=5 \times 10^{-6}$, (f) $t=6 \times 10^{-6}$, (g) $t=7 \times 10^{-6}$, (h) $t=8 \times 10^{-6}$.}
	\label{figs:HVI-stress}
\end{figure}

\begin{figure}[htb!]
	\centering
	\includegraphics[trim = 0cm 0cm 0cm 0cm, clip,width=1\textwidth]{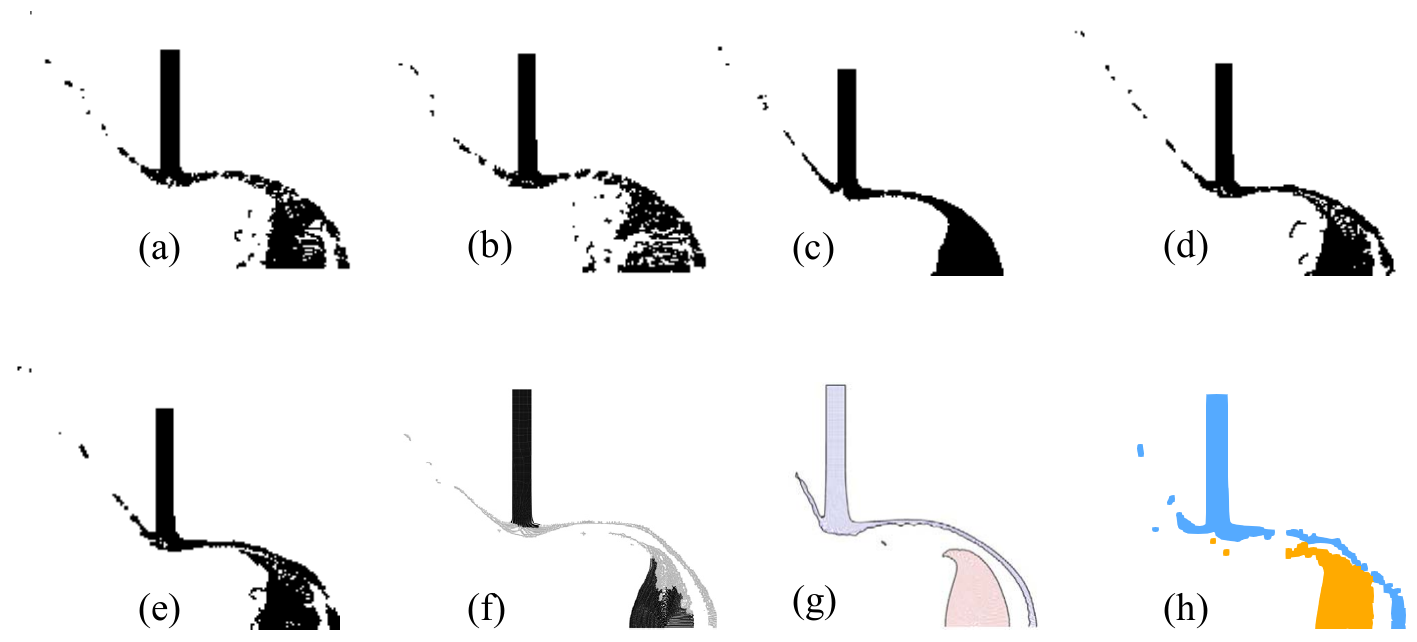}
	\caption{High-velocity impact: deformation of projectile and target at $t=8 \times 10^{-6}$ obtained with different algorithms, i.e., (a) BAL \cite{mehra2006high}, (b) MON \cite{mehra2006high}, (c) CON \cite{mehra2006high}, (d) SAV1 \cite{mehra2006high}, (e) SAV2 \cite{mehra2006high}, (f) adaptive total Lagrangian Eulerian SPH \cite{young2021adaptive}, (g) Xiao and Liu's study \cite{xiao2023penalty}, (h) the present SPH-GNOG.}
	\label{figs:HVI-compare}
\end{figure}

The deformation of the projectile and target at different time points is illustrated in Fig. \ref{figs:HVI-stress}, colored according to von Mises stress. 
Throughout the simulation, no zigzag distribution of particles and stresses, in other words, no occurrence of hourglass modes, are observed.
Then the upper half of the particle configure at time $t=8 \times 10^{-6}$ is compare with previous numerical results, as shown in Fig. \ref{figs:HVI-compare}.
Mehra and Chaturvedi \cite{mehra2006high} addressed this problem using five variations of SPH, referred to as BAL, MON, CON, SAV1, and SAV2.
BAL refers to a scheme that incorporates the Balsara switch \cite{balsara1995neumann}, designed to limit excessive artificial viscosity. MON involves a modification to artificial viscosity introduced by Morris and Monaghan \cite{morris1997switch}. CON represents an SPH formulation that solves the conservation equations by applying a solution to the Riemann problem \cite{parshikov2002smoothed, parshikov2000improvements}.
SAV1 and SAV2 correspond to conventional Eulerian SPH with artificial viscosity parameters \cite{gingold1977smoothed} set at $\alpha =1, \beta =2$ and $\alpha =2.5, \beta =2.5$, respectively.
The results from an adaptive total Lagrangian Eulerian SPH \cite{young2021adaptive} and Xiao and Liu's penalty-based surface-to-surface contact SPH algorithm \cite{xiao2023penalty} are also shown for comparison.
It can be observed that the result obtained using the present SPH-GNOG is similar to previous results in terms of deformation and particle distribution at $t=8 \times 10^{-6}$, further validating the stability and accuracy of the proposed algorithm.

\subsection{Striker pin fuse}
\label{pin-fuse}

A striker pin fuse is a device designed to protect electrical circuits from overload \cite{jacobs1963electric}. 
When an excessive current is detected, the spring-loaded striker pin is released and propelled towards the shear plate. 
Upon impact, the shear plate fractures, thereby interrupting the circuit to prevent further damage.
The shear plate is designed with pre-defined weak points that serve as fracture lines, ensuring the circuit can be promptly interrupted when necessary.
This section simulates the process in which the striker pin strikes the shear plate, leading to its rupture, in order to demonstrate the potential industrial applications of the proposed SPH-GNOG.

\begin{figure}[htb!]
	\centering
	\includegraphics[trim = 0cm 0cm 0cm 0cm, clip,width=0.5\textwidth]{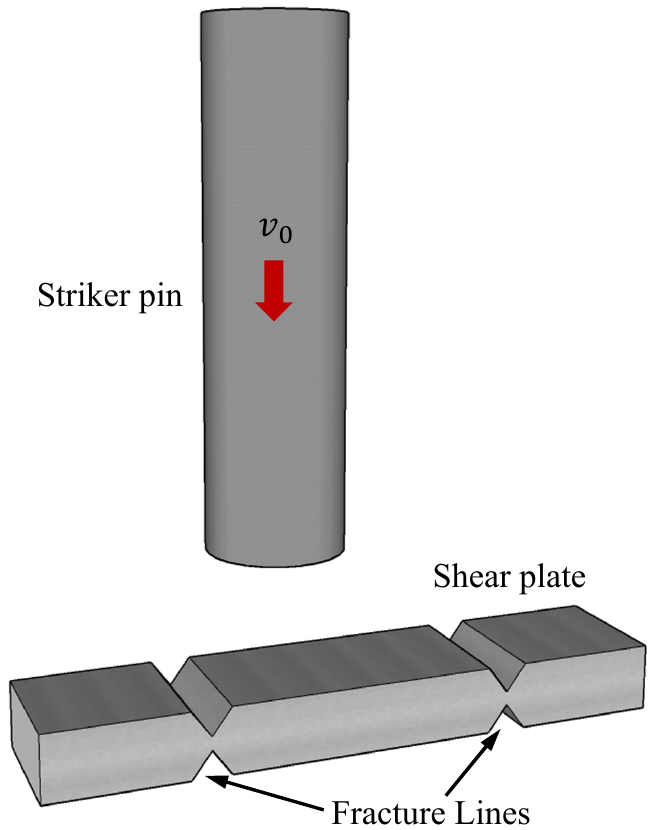}
	\caption{Striker pin fuse: model setup.}
	\label{figs:fuse-setup}
\end{figure}

Fig. \ref{figs:fuse-setup} presents a simplified model setup of the striker pin and shear plate. The striker pin, made of steel, is cylindrical with a radius of $r=0.002$ and a length of 0.015. The shear plate, constructed from aluminum, is fixed at both ends and has dimensions of 0.02 in length, 0.004 in width, and 0.002 in thickness. 
Both components are simulated using the $J_2$ plasticity model with linear hardening law. The material parameters are detailed in Table \ref{fuse-parameter}. The initial particle spacing is set to $dp=r/15$ and the initial velocity $v_0=40$.
A failure model is employed in which a particle will fail if the negative pressure (tensile stress) exceeds the threshold $p_{min}$ \cite{ma2009comparison}. 
When failure occurs, pressure values will no longer be allowed to go negative in subsequent computations \cite{ma2009comparison}.
The value of $p_{min}$ is set to $-8 \times 10^{8}$ \cite{young2021adaptive} for the shear plate and the damage is not considered for the striker pin.

\begin{table}
	\scriptsize
	\centering
	\caption{Striker pin fuse: material properties of the striker pin and the shear plate.}
	\begin{tabularx}{8cm}{@{\extracolsep{\fill}}lcccccc}
		\hline
		$ \ $ & $\rho$ & $E \ [\times 10^{9}]$ & $\nu$ & $\sigma_Y \ [\times 10^{6}]$ & $\kappa \ [\times 10^{6}]$ \\
		\hline
        Striker pin & 7790 & 193 & 0.33 & 566 & 832.8 \\
		Shear plate & 2780 & 71 & 0.31 & 290 & 456.4 \\
		\hline
	\end{tabularx}
	\label{fuse-parameter}
\end{table}

\begin{figure}[htb!]
	\centering
	\includegraphics[trim = 0cm 0cm 0cm 0cm, clip,width=1\textwidth]{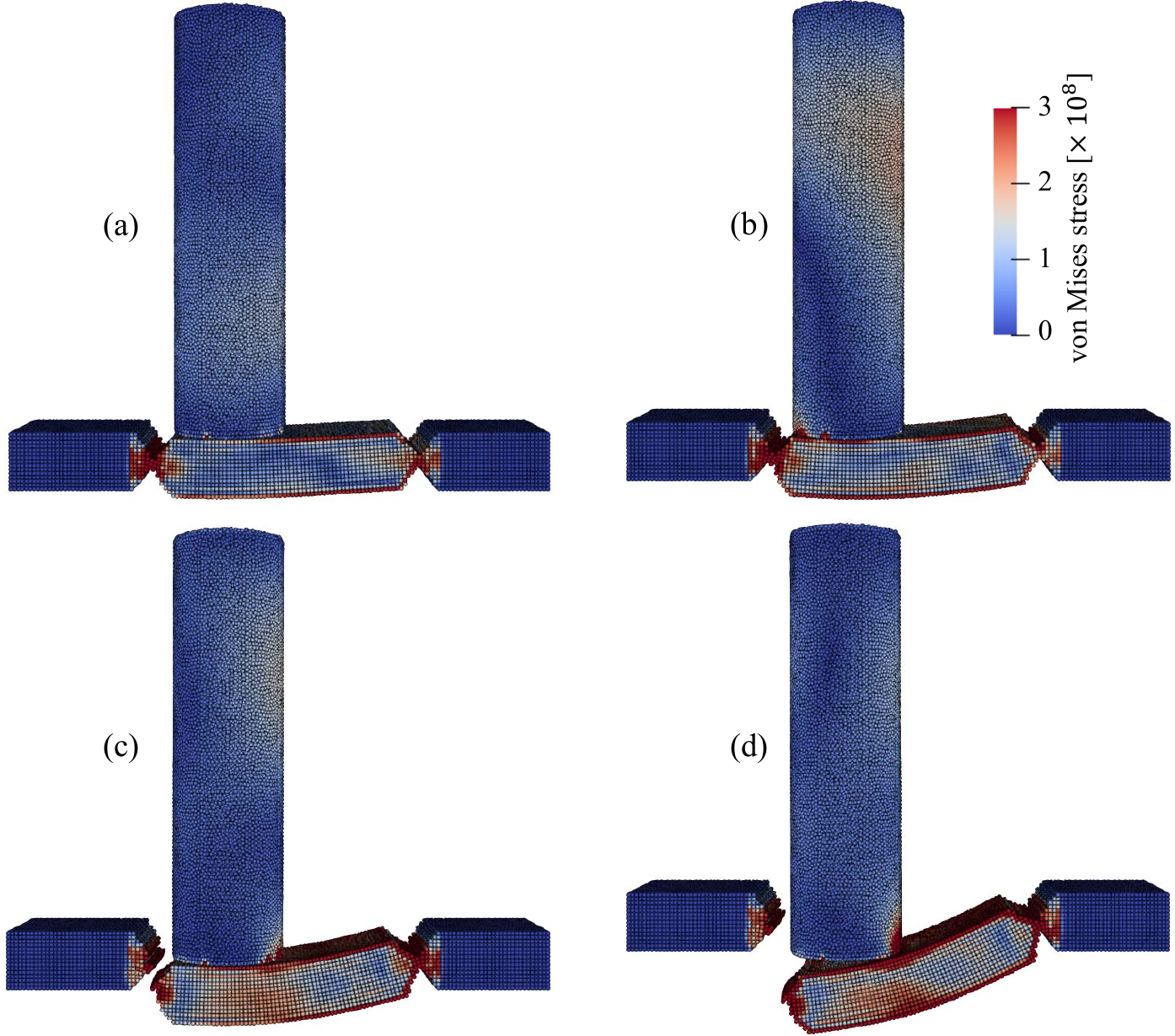}
	\caption{Striker pin fuse: evolution of von Mises strain with the present SPH-GNOG at (a) $t=1 \times 10^{-5}$, (b) $t=2 \times 10^{-5}$, (c) $t=4 \times 10^{-5}$, (d) $t=1 \times 10^{-4}$.}
	\label{figs:fuse-stress}
\end{figure}

Fig. \ref{figs:fuse-stress} illustrates the particle and von Mises stress distribution for the striker pin and shear plate at various time intervals, with time zero defined as the moment when the striker pin first makes contact with the shear plate. 
When the striker pin impacts the metal plate, stress concentration occurs at the predefined weak points, causing the plate to fracture along these lines, thereby effectively severing the circuit.
During the simulation, the particle distribution remains uniform, and the stress distribution appears smooth, with no observed hourglass modes. 
This indicates the potential of the proposed formulation for applications in industrial scenarios.
%
%
\section{Conclusions}
\label{conclusions}
This study develops a generalized non-hourglass formulation for ULSPH, applicable to both elastic and plastic materials (including perfect plasticity and linear hardening plasticity), by introducing a penalty force to resolve inconsistencies between the predicted linear velocity and the actual velocity of neighboring particle pairs. 
Notably, this approach eliminates the need for case-dependent parameter tuning.
Through comprehensive validation using benchmark cases and the incorporation of a dual-criteria time-stepping scheme to enhance computational efficiency, we demonstrate the robustness and accuracy of the proposed SPH-GNOG method. 
Furthermore, the potential industrial applications of the formulation are highlighted in Section \ref{pin-fuse}.
The key advantages of the proposed method are as follows:

(1) The performance of the current SPH-GNOG in classical elastic scenarios, such as the oscillating plate and colliding rubber ring, is comparable to that of SPH-ENOG \cite{zhang2024essentially} and remarkable surpasses that of SPH-OG and SPH-OAS \cite{gray2001sph}.

(2) In scenarios where angular momentum plays a significant role, such as the spinning plate, the present SPH-GNOG significantly outperforms SPH-ENOG \cite{zhang2024essentially}.

(3) The proposed method is applicable to plastic materials, specifically for high-speed impact and failure of metallic materials.
%
%
\section*{CRediT authorship contribution statement}
\addcontentsline{toc}{section}{CRediT}

Shuaihao Zhang: Conceptualization, Methodology, Investigation, Visualization, Validation, Formal analysis, Writing - original draft, Writing - review \& editing. 
Dong Wu: Investigation, Methodology, Formal analysis, Writing - review \& editing. 
Sérgio D.N. Lourenço: Supervision, Writing - review \& editing. 
Xiangyu Hu: Supervision, Investigation, Methodology, Writing - review \& editing.
%
%
\section*{Declaration of competing interest}
\addcontentsline{toc}{section}{declaration-interest}

The authors declare that they have no known competing financial interests or personal relationships that could
have appeared to influence the work reported in this paper.

%
%
\section*{Data availability}
\addcontentsline{toc}{section}{data-availability}

The code and data are available as open-source through the SPHinXsys project: \href{https://www.sphinxsys.org}{https://www.sphinxsys.org}.

%
%
\section*{Acknowledgements}
\addcontentsline{toc}{section}{acknowledgement}

Dong Wu and Xiangyu Hu would like to express their gratitude to the German Research Foundation (DFG) for their sponsorship of this research under grant number DFG HU1527/12-4.
Sérgio D.N. Lourenço would like to express his gratitude to the Research Grants Council Hong Kong for their sponsorship of this research under a Collaborative Research Fund (C6006-20GF).
The computations were performed using research computing facilities offered by Information Technology Services, the University of Hong Kong.
%
%

%
%
\bibliographystyle{elsarticle-num}
\bibliography{hourglass-control}
%
%
\end{document}